\documentclass[preprint]{acmsiggraph}

\TOGonlineid{0000}

\TOGvolume{}
\TOGnumber{}

\TOGarticleDOI{}

\TOGprojectURL{}
\TOGvideoURL{}
\TOGdataURL{}
\TOGcodeURL{}

\title{Rigorous asymptotic and moment-preserving diffusion approximations for generalized linear Boltzmann transport in \changed{arbitrary} dimension}

\author{ Eugene d'Eon \\
Jig Lab, Wellington, New Zealand }

\pdfauthor{ Eugene d'Eon }


\usepackage{smrdefaults}

\usepackage{color}

\usepackage{multirow}

\usepackage{rotating}
\newcommand{\changed}[1]{#1}
\newcommand{\changedcomment}[2]{#2}

\newcommand{\F}{\mathcal{F}_d}  
\renewcommand{\Finv}{\mathcal{F}_d^{-1}}  

\newcommand{\ssalbedo}{c}

\renewcommand{\o}{\overline}

\newcommand{\phipt}{\phi_{\text{pt}}}  
\newcommand{\Cpt}{C_{\text{pt}}}  
\newcommand{\Qpt}{Q_{\text{pt}}}  

\newcommand{\vect}[1]{\vec{#1}}

\newcommand{\e}{\text{e}}

\newcommand{\n}{\vect{\mathbf{n}}}

\newcommand{\dir}{\vect{\omega}}

\def\rm{R^{-}}

\usepackage{paralist}
\usepackage{subfigure}
\usepackage{rotating}
\usepackage{amsgen}

\usepackage{alltt}

\begin{document}




\maketitle


\changedcomment{Dear reviewers: Thank you for your reviews.  Changes to the manuscript are colored blue.  Comments are sometimes provided [bold, in square brackets] prior to the changes to describe what changed and why.
Regarding the size of Figure 2: agreed - I would expect that upon acceptance the paper would be formatted carefully using the typesetting style of the journal (thus, the figure remains unadjusted for now).
Regarding Figure 8b: The feature you noticed is not a statistical phenomena: the Pearson random flight has a Dirac delta free-path distribution and every scattering event is followed by a deterministic (in this case, unit) step length.  Thus, each order of scattering has a finite support and the sum of the various scattering orders produces a non-smooth curve.  This form of random walk was considered along with the rest because it, in a sense, provides the most extremely correlated walk one might imagine (despite it being non-physical and not corresponding to any particular medium that I can think of).  Thus, in considering a variety of uncorrelated, and correlated transport models, we decided to consider this extreme case (and additionally, because it is specifically studied in mathematics).  Despite the discontinuous form of the exact solution, it was interesting to see how the various (smooth) diffusion approximations behave.  We have included additional comments to make this more clear to the reader.  We were not able to numerically invert the exact solution (involving three Bessel functions) in a stable way, so we chose to include Monte Carlo results as the benchmark solution.}{}

\begin{abstract}
  We derive new diffusion solutions to the monoenergetic generalized linear Boltzmann transport equation (GLBE) for the stationary collision density and scalar flux about an isotropic point source in an infinite $d$-dimensional absorbing medium with isotropic scattering.  We consider both classical transport theory with exponentially-distributed free paths in arbitrary dimensions as well as a number of non-classical transport theories (non-exponential random flights) that describe a broader class of transport processes within partially-correlated random media.  New rigorous asymptotic diffusion approximations are derived where possible.  We also generalize Grosjean's moment-preserving approach of separating the first (or uncollided) distribution from the collided portion and approximating only the latter using diffusion.  We find that for any spatial dimension and for many free-path distributions Grosjean's approach produces compact, analytic approximations that are, overall, more accurate for high absorption and for small source-detector separations than either $P_1$ diffusion or rigorous asymptotic diffusion.  These diffusion-based approximations are exact in the first two even spatial moments, which we derive explicitly for various non-classical transport types.  We also discuss connections between the random-flight-theory derivation of the Green's function and the discrete spectrum of the transport operator and report some new observations regarding the discrete eigenvalues of the transport operator for general dimensions and free-path distributions.
\end{abstract}




\keywordlist






\section{Introduction}

  Predicting the bulk transport of particles or waves within random media is a long-studied problem in statistical physics~\cite{chandrasekhar43,williams71}.  Applications of such analysis include the design of nuclear weapons and reactors~\cite{weinberg58,bell70}, the study of light distributions in stellar media and tissue~\cite{chandrasekhar60,ishimaru78,tuchin07}, remote atmospheric sensing~\cite{marshak05}, radiative heat transfer~\cite{modest03} and computer graphics~\cite{jakob10,deon11a}.  A common approximate mathematical framework at the core of all of these applications is linear transport theory, known to be intimately connected to the Pearson theory of random flights~\cite{grosjean53,guth60}.  With very few exceptions, exact solutions to simple ``model" problems have complicated integral forms that are costly to evaluate.  More general ``real-world" problems typically have no known exact solutions, placing great importance on efficient and accurate approximate methods.

  Diffusion theory provides a more aggressively approximate framework for studying many bulk transport processes~\cite{rossum99} and is often accurate within some subdomain of a given problem.  \changed{Solutions $\phi$ of the \emph{diffusion equation}
  \begin{equation}
    -D \, \nabla^2 \phi(\vec{x}) + \Sigma_a \phi(\vec{x}) = Q(\vec{x})
  \end{equation}
  and more readily obtainable, either exactly or approximately, and are used to estimate part or all of some quantity of interest in the original transport problem, where $D$ and $\Sigma_a$ are typically constants and $Q$ describes the initial source of particles.}

  \changed{There are a variety of different ways to relate the diffusion equation to a given transport problem.  Despite a} rampant widespread conclusion that there is only \emph{one} diffusion approximation and that it is necessarily inaccurate for high absorption levels and for small source-detector separations, \changed{there are, in fact}, a variety of proposed modified diffusion theories and, when chosen carefully, these theories can provide improved accuracy \changed{at} negligible extra cost.  No one diffusion-based approximation is universally superior for all linear transport problems, so it is of interest to evaluate non-classical diffusion in light of recent developments in transport theory.  In this paper we revisit two specific forms of ``non-classical" diffusion approximations:
  \begin{compactitem}
    \item An asymptotic ``rigorous" diffusion approximation formed by discarding all transient terms corresponding to the continuous portion of the spectrum of the transport operator (leaving only the discrete terms)~\cite{case67} 
    \item A moment-preserving ``uncollided plus diffusion" approximation by Grosjean~\shortcite{grosjean56a} that decouples diffusion from uncollided flux and performs well, especially for high absorption levels and for small source-detector separations, where most other diffusion approximations do not.
  \end{compactitem}
  Both of these variations of diffusion theory complement the \emph{parameter-of-smallness} asymptotic approximations derived by Larsen and Vasques\changedcomment{adding important Larsen 2010 citation}{}~\cite{larsen10,larsen11} as well as the $P_1$ (or, perhaps, \emph{classical}) diffusion approximation (which can be derived in several ways).  The proposal of alternative (non-$P_1$) diffusion for improving accuracy for high absorption levels began with Wigner~\shortcite{wigner43} in neutron transport, \changedcomment{expanding slightly}{followed by Grosjean~\shortcite{grosjean54,grosjean56a,grosjean56b,grosjean63b}.  Applications in other fields include several investigations in tissue optics~\cite{arnfield92,kim98,graaff01} and Grosjean's modified diffusion theory} sees continued used in film visual effects since its introduction to computer graphics~\cite{deon11a}.

  Our first generalization of previous diffusion techniques is to consider arbitrary dimension $d \ge 1$ of the space in which the transport occurs.  Grosjean's form of diffusion approximation has only been derived and evaluated for three-dimensional domains.  It is not purely a matter of mathematical interest to consider transport in various dimensions.  Transport in a one-dimensional domain (the \emph{rod model}~\cite{wing62}, also known as the Fermi model or the forward/backward model) has many application including transmission line theory~\cite{redheffer62}, for constructing the two-stream approximation in many three-dimensional problems~\cite{peraiah02} for velocity jump processes in biology~\cite{othmer00} and for teaching and discovering new deterministic and Monte Carlo techniques~\cite{hoogenboom08a}.  For \emph{classical} transport in one dimension, diffusion happens to be an exact solution.  This is not always true in the rod-model, and later in the paper we show how to derive free-path distributions such that diffusion is an exact solution in higher dimensions.   Additionally, two-dimensional linear transport theory is useful for describing search strategies and predicting the migration of animals and diseases~\cite{yang09}, for describing transport along thin conductive surfaces~\cite{bal00b}, for describing scattering in three-dimensional media with highly anisotropic structures such as fibrous tissue, aligned hair clusters, wood and ice~\cite{meylan06} and also within thin films~\cite{vynck12}.  Two-dimensional transport solutions also arise in approximate three-dimensional theories such as ``weakly" three-dimensional Caseology~\cite{pomraning96}.  It is also interesting in general to study the effect of spatial dimension on various aspects of transport theory, such as the spectrum of the transport operator. We are unaware of any study of the discrete spectrum of the transport operator (and, thus, rigorous diffusion) in any dimensions other than three or one.    

  The second generalization of previous diffusion techniques that we consider is to permit random flights with non-exponential free path distributions.  Classical transport theory assumes infinitesimal scattering centers with completely uncorrelated positions, leading to a free-path distribution (and extinction law) that is exponential.  A \emph{generalized linear Boltzmann equation} (GLBE) has recently been formalized~\cite{larsen11,vasques13}, permitting general free-path distributions.  This widens the scope of transport theory to encompass a much broader class of random media.  Recent studies of such non-classical transport include light transport in partially-correlated random media~\cite{kostinski01,davis04,moon07} and also for neutron transport---especially for pebble-bed reactors~\cite{vasques09,vasques13,vasques13b}.  Random flight theory has long considered arbitrary free path distributions, but, with the exception of Grosjean~\shortcite{grosjean51,grosjean53}, who studied infinite medium transport problems in complete generality, rather little non-exponential transport theory has been studied until fairly recently.  To study the relationship between free-path distribution and diffusion theory, we have selected a variety of parametric non-exponential free path distributions and derived diffusion approximations for each of them in arbitrary dimension.  We mention results for beta-prime-, chi-, gamma- and delta- (Pearson-) random flights.  None of these chosen distributions describe any specific correlated random media that we are aware of, but we have considered a rather wide variety of distribution shapes that we expect will find application, if only approximate, in future work.

  We remark that the GLBE constitutes a new \emph{generalized volume rendering equation} for computer graphics, permitting image synthesis of participating media that violate the assumptions of classical transport theory.  This is topical, given recent claims that no physical media of any kind has exactly exponential free path distributions~\cite{davis04}.  The GLBE should also provide a powerful level-of-detail tool for compressing explicit optical complexity with a statistical model that accurately reproduces the bulk behaviour of light transport within partially-correlated collections of material.  Preliminary work along these lines has already been explored in graphics~\cite{moon07}.  Application of the GLBE to Monte Carlo rendering algorithms and for deriving new variance reduction techniques for estimators in generalized media is straightforward~\cite{larsen11}.  Our new diffusion approximations widen the potential application of the GLBE to rendering more general media by extending previous diffusion-based analytic tools based on the method of images, which are highly popular in medical physics and computer graphics~\cite{farrell92,jensen01,donner05,deon11a,habel13,deon13a}, as well as finite element diffusion methods and their hybrids~\cite{ackroyd81,arbree11}.

  \subsection{Scope and related work}
    Our investigations never stray far from the classic problem of an isotropic point source in an infinite homogeneous medium with isotropic scattering and absorption.  We chose the point source problem for its close connection to both linear transport theory and random flight theory.  Diffusion asymptotics for the isotropic point source transfer directly to plane-parallel problems, such as the half-space and slab geometry, as well as searchlight problems, where the same diffusion lengths appear, so what may appear to be a narrow focus nevertheless yields rather broad reaching results.  We consider only isotropic scattering in this paper.  For exact Fourier-inversion solutions of the time-dependent and time-independent point source and plane source infinite medium problems in three dimensions with arbitrary anisotropic scattering and arbitrary free-path distributions, see Grosjean's monograph~\shortcite{grosjean51}.  Likewise, for the time-independent point source $d$-dimensional infinite medium problems with anisotropic scattering and general free-path distribution, see~\cite{grosjean53} and~\cite{degregorio12}.

  \subsection{Outline}
    In the next section we begin by recalling the Fourier transform approach for solving transport problems with spherical symmetry.  Familiar exact and diffusion solutions for the isotropic point source problem in three dimensional classical transport are included alongside the presentation of the general theory.  \changedcomment{More detail to prepare the reader for a laundry list of solutions}{  In Section~\ref{sec:rigorous} we recall the rigorous asymptotic diffusion approximation, whose form we extract directly from poles of the Fourier-transformed exact solutions.  Exact moments of the scalar flux and fluence are also derived directly from the Fourier transformed solutions in Section~\ref{sec:moments} and these provide the basis for deriving Grosjean's modified approach to diffusion approximations, in Section~\ref{sec:grosjean}.  Here we also conjecture a simplified method for deriving Grosjean's diffusion coefficients given only the mean free path and mean square free path of the medium.  This conjecture is supported in the remaining sections, which consider transport processes in spaces with a variable number of dimensions under a variety of free path distributions.  Our findings from this generalization study are reported in the remaining sections, beginning} in Section~\ref{sec:exponential} with classical exponential transport with spatial dimensionalities other than three.  A variety of non-exponential generalized transport models are then discussed in the subsequent sections.  \changed{These sections are mostly comprised of equations detailing any exact and diffusion solutions that we were able to find, with some limited discussion and comparisons along the way.  We summarize our findings in the final sections.}
\section{The isotropic point source Green's function for the infinite medium}

  We seek to predict the angular flux/intensity/radiance $L(\vec{r},\dir)$ of particles at any position $\vec{r}$ in direction $\dir$ within an infinite homogeneous medium.  Specifically, we consider the classic problem of an isotropic point source at the origin of an infinite $d$-dimensional homogeneous medium with absorption and isotropic scattering.  Particles undergo a random flight leaving the origin in an isotropically random direction and move a distance $s$ sampled from the \emph{free path distribution} $p(s)$, a normalized probability distributions on $[0,\infty)$:
  \begin{equation}
  	\int_0^\infty p(s) ds = 1.
  \end{equation}
  The particles suffer \emph{collision} events after each displacement.  At each collision, with probability $\ssalbedo$, a particle chooses a new isotropic random direction and takes another step.  With probability $1-\ssalbedo$, the random flight is terminated (the particle is absorbed).  In radiative transfer, the quantity $\ssalbedo$ is often referred to as the \emph{single-scattering albedo} of the medium (or the average number of secondaries per collision, in neutron transport).

  We begin with a quantity at the heart of random flight theory, the \emph{$n$th collision densities} $\Cpt(r|n)$---the probability densities for particles to \emph{enter} their $n$th collision at a distance $r$ from the point source.  This quantity is useful for saying something about the behaviour of a single particle.  In transport theory, however, we typically seek only the reaction rates or total densities of many particles at some position and direction and no information about any individual particle's previous events is known or relevant.  However, the random flight and transport quantities are closely related, the latter being the sum of the former.  The \emph{total collision density} $\Cpt(r)$ is
  \begin{equation}
  	\Cpt(r) = \sum_{n=1}^\infty \Cpt(r|n).
  \end{equation}
  To acquire the angular flux / radiance, we need to know the densities of particles \emph{leaving} collisions, as well as the \emph{extinction function} $E(s)$ for the medium,
  \begin{equation}
    E(s) = 1 - \int_0^s p(s') ds'.
  \end{equation}
  We refer to the distribution of particles \emph{leaving} their $n$th collision at a distance $r$ from the point source as the $n$th \emph{collision source density} $\Qpt(r|n)$, which is simply
  \begin{equation}
  	\Qpt(r|n) = \ssalbedo \, \Cpt(r|n).
  \end{equation}
  The point source together with the total density of particles leaving previous collisions $\Qpt(r)$
  \begin{equation}
  	Q(r) = \delta(r) + \Qpt(r) = \delta(r) + \sum_{n=1}^\infty \Qpt(r|n)
  \end{equation}
  constitutes the \emph{effective source density} $Q(r)$ within the medium.  We can then relate $Q(r)$ to the transport angular flux / fluence.  In the case of isotropic scattering in 3D, the angular flux / radiance at radius $r$ from point source and in direction $\mu$ (where $\mu$ is the cosine of the angle between the direction $\dir$ and the normalized position vector) can be compactly expressed as a line integral of the effective source density~\cite{davison00},
  \begin{equation}
    L(r,\mu) = \int_0^\infty \Qpt(\sqrt{R^2+r^2-2 r R \mu}) E(R) dR,
  \end{equation}
  which includes the singular term
  \begin{equation}
    \frac{1}{4 \pi r^2} E(r) \delta(\mu - 1)
  \end{equation}
  (by including $\delta(r)$ in $Q(r)$).
  
  Another quantity of interest in transport problems is the \emph{scalar flux} or \emph{fluence} $\phipt(r)$, which is proportional to the density of particles \emph{in flight} at any position in the medium and is not to be confused with the collision density $\Cpt(r)$.  The scalar flux is the integral of the angular flux / radiance over all directions and can similarly be written as the Neumann sum of \emph{$n$th-collided scalar fluxes} $\phipt(r|n)$:
  \begin{equation}
  	\phipt(r) = \sum_{n=0}^\infty \phipt(r|n).
  \end{equation}
  These $n$th-collided scalar fluxes $\phipt(r|n)$ are proportional to the density of particles \emph{in flight} at some radius $r$ from the point source that have experienced exactly $n$ previous scattering events on their way from the source to a radius $r$.  The scalar flux for $n = 0$ is referred to as the \emph{uncollided flux} $\phipt(n|0)$.  The $n$th-collided scalar flux is the convolution of $\Qpt(r|n)$ with a distribution related to the extinction function $E(s)$.
  The total scalar flux / fluence of the medium is similarly the convolution of the effective source distribution with a distribution involving the extinction function---a relationship that we formalize in the next section.

  A clear distinction between collision density $\Cpt$ and scalar flux $\phipt$ becomes essential for generalized Boltzmann transport.  Only when the free path distribution $p(s)$ is exponential are the collision density and scalar flux proportional~\cite{larsen11}.  This aspect of the GLBE creates a new pair of distinct measurable quantities that might have previously appeared to be the same.  We will show later that the non-exponential free-path distributions of the GLBE also lead to pairs of distinct diffusion approximations, one for the collision density and one for the scalar flux / fluence.  The collision density may be of interest in problems such as radiation therapy, dosimetry problems, or when the heat created by absorption events is of interest---i.e. anytime the medium itself is doing the measurement, so to speak (when detection at some position is dependent on the step distance $s$ since last collision/birth).  On the other hand, the scalar flux is more likely of interest for predicting the quantity measured by some small detector placed in the medium, such as a camera in a foggy atmosphere (ie. when measurement is not dependent on $s$).  This distinction, of course, only arises for measurements internal to the scattering volume.  Measurements of exitant radiance or exitant flux at a boundary are the only measurable quantity, regardless of free path distribution.

  Without loss of generality we simplify our derivations by assuming a unit of time such that the particle velocity is unity.  For much of the presentation we also assume a unit of distance such that the \emph{mean-free path} $\ell$ (regardless of free-path distribution function) is also unity.  The problem is then characterized only by the spatial dimension $d$, the single-scattering albedo $\ssalbedo$, and the free-path distribution, $p(s)$ satisfying
  \begin{equation}
  	\ell = \int_0^\infty s p(s) ds = 1.
  \end{equation}
  We assume isotropic scattering at each scattering event and assume that the free path distribution is the same for all orders of scattering (more general transport problems with arbitrary free path distributions $p_n(s)$ at each step have been considered by Grosjean~\shortcite{grosjean51,grosjean53}).  We also assume that absorption arises due to interaction with the same particles responsible for collision (and scattering) (or, equivalently, that the free-path distribution for interacting with absorbing particles is identical to the free-path distribution for interacting with scattering particles).  Therefore the single-scattering albedo $c$ remains a constant, independent of $s$.

  \subsection{Solution via Fourier transform}
  	We use a Fourier transform approach~\cite{grosjean51,grosjean53,zoia11e} to solve for the collision densities, scalar fluxes and the moments of both.  Thus, the following derivation is largely in the style of random flight theory~\cite{dutka85}, but produces the well known solution that satisfies the transport equation.

  	Particles leave the point source in a uniform random direction and move a random distance sampled from the free-path distribution function $p(s)$.  This produces the \emph{uncollided propagator} $\zeta_d(r)$ defined as
  	\begin{equation}
  		\zeta_d(r) = \frac{p(r)}{\Omega_d(r)}
  	\end{equation}
  	where
  	\begin{equation}
  		\Omega_d(r) = \frac{d r^{d-1} \pi^{d/2}}{\Gamma(d/2+1)}
  	\end{equation}
  	is the surface area of an $d$-dimensional sphere with radius $r$ and $\Gamma$ is the Euler gamma function.  The uncollided propagator $\zeta_d(r)$ gives the probability distribution of displacements $r$ for a single step of the random flight.

  	The random flight / transport process is formed iteratively by repeated application of the uncollided propagator, interleaved with absorption.  Due to the spherical symmetry of the problem (given both isotropic emission and scattering), this source iteration can be expressed compactly using convolution, starting with the point source initial condition:
  	\begin{equation}\label{initialcond}
  		\Qpt(r|0) = \delta(r)
  	\end{equation}
  	and relating the density of collisions of order $n+1$ to those of order $n$ via~\cite{grosjean51}
  	\begin{equation}
  		\Cpt(r|n+1) = \int \zeta_d(|\vec{r}-\vec{r'}|) \Qpt(r'|n) d\vec{r}'.
  	\end{equation}
  	The convolutions are easily expressed using spherically-symmetric Fourier transforms~\cite{grosjean53,dutka85}.  The forward and inverse spherical Fourier transforms of a radially-symmetric function $f(r)$ (with radius $r \ge 0$) in the general $d$-dimensional case are
  	\begin{align}\label{eq:fouriertransforms}
  		\F\{f(r)\} = z^{1-d/2} (2\pi)^{d/2} \int_0^\infty r^{d/2} J_{d/2-1}(r z) f(r) dr. \\
  		\Finv\{\o{f}(z)\} = r^{1-d/2} (2\pi)^{-d/2} \int_0^\infty z^{d/2} J_{d/2-1}(r z) \o{f}(z) dz
  	\end{align}
  	where $z$ is the transformed coordinate relating to $r$ and $J_k$ is the modified Bessel function of the first kind.  Note that these transforms are well defined for arbitrary (even non-integer) dimensions $d \ge 1$.

  	The transformed collision density of order $n$ is $\o{\Cpt}(z|n) = \mathcal{F_d}\{ \Cpt(r|n) \}$, and, by repeated convolution in frequency space, results in
  	\begin{equation}\label{eq:Czn}
  		\o{\Cpt}(z|n) = \ssalbedo^{n-1} \left( \o{\zeta}_d(z) \right)^n,
  	\end{equation}
  	which requires the \emph{transformed free propagator}
    \begin{equation}\label{eq:zetaz}
      \o{\zeta}_d(z) = \F\{ \zeta_d(r) \}.
    \end{equation}  
    Likewise the transformed $n$th collision source density is
  	\begin{equation}
  		\o{\Qpt}(z|n) = \ssalbedo^{n} \left( \o{\zeta}_d(z) \right)^n.
  	\end{equation}
  	We have assumed a non-multiplying medium with non-zero absorption ($0 < \ssalbedo < 1$), so the transformed total collision density of all orders can be expressed as the limit of the geometric series,
  	\begin{equation}\label{eq:Cz}
  		\o{\Cpt}(z) = \frac{\o{\zeta}_d(z)}{1-\ssalbedo \o{\zeta}_d(z)},
  	\end{equation}
  	and, likewise,
  	\begin{equation}
  		\o{\Qpt}(z) = \frac{\ssalbedo \, \o{\zeta}_d(z)}{1-\ssalbedo \o{\zeta}_d(z)}
  	\end{equation}
  	
    The scalar fluxes $\phi(r)$ are found by convolving the collision source densities with the \emph{stretched extinction} function $X(r)$,
    \begin{equation}
      X(r) = \frac{E(r)}{\Omega_r(r)}.
    \end{equation}
    The stretched extinction function is much like the collision propagator, but corresponds to the density of particles that have moved uncollided along a distance $r$ since their last collision/birth (as opposed to the fraction of those that then create the density of collisions at $r$).
  	Performing the required convolution in frequency space produces the transformed \emph{uncollided scalar flux}
  	\begin{equation}
  		\o{\phipt}(z|0) = \o{X}(z),
  	\end{equation}
  	the transformed scalar flux for particles that have experienced exactly $n$ collisions
  	\begin{equation}\label{eq:phizn}
  		\o{\phipt}(z|n) = \o{X}(z) \left( \ssalbedo \o{\zeta}_d(z) \right)^n,
  	\end{equation}
  	and the total transformed scalar flux
  	\begin{equation}\label{eq:phizunstable}
  		\o{\phipt}(z) = \frac{\o{X}(z)}{1-\ssalbedo \o{\zeta}_d(z)}.
  	\end{equation}
  	The subject of this paper is to derive exact results (or accurate approximations) for the inverse transforms of these Fourier quantities to produce the $n$th and total collision densities and scalar fluxes / fluences for the isotropic point source problem.  Often these transformed distributions are not analytically invertible and numerical inversion is required.  For convergence of the inverse transform it is necessary~\cite{grosjean63b} to separate the uncollided distribution from the scattered distribution,
  	\begin{equation}\label{eq:Czcollided}
  		\o{\Cpt}(z) = \o{\zeta}_d(z) + \frac{\ssalbedo \o{\zeta}_d(z)^2}{1-\ssalbedo \o{\zeta}_d(z)},
  	\end{equation}
  	\begin{equation}\label{eq:phiz}
  		\o{\phipt}(z) = \o{X}(z) + \frac{\o{X}(z) \ssalbedo \o{\zeta}_d(z) }{1-\ssalbedo \o{\zeta}_d(z)}.
  	\end{equation}
  	  This section has recalled well-known random flight theory that has been applied previously to general transport problems~\cite{grosjean51,grosjean53,zoia11e}.  Next, we first recall the application of this theory to solve the classic 3D point source problem before deriving new diffusion approximations.

  \subsection{Classical transport theory in three-dimensions}
  	Classical transport theory assumes an exponentially-distributed free-path distribution with mean free path $\ell$,
  	\begin{equation}
  		p(s) = \frac{1}{\ell}\e^{-s / \ell}
  	\end{equation}
  	The extinction is also exponential, $E(s) = \e^{-s / \ell}$.  The exponential free-path distribution is the only free-path distribution such that the \emph{interaction coefficient} $\Sigma_t$
    \begin{equation}
      \Sigma_t(s) = \frac{p(s)}{E(s)} = \frac{1}{\ell}
    \end{equation}
    is independent of $s$~\cite{larsen11}.  In three dimensions ($d = 3$), the uncollided propagator simplifies to
  	\begin{equation}
  		\zeta_3(r) = \frac{\e^{-r / \ell}}{4 \pi r^2}
  	\end{equation}
  	and the Fourier-transformed propagator is then~\cite{placzek43}
  	\begin{equation}
  		\o{\zeta}_3(z) = \frac{1}{z} \int_0^\infty \frac{p(r) \sin(r z)}{r} dr = \frac{\arctan (\ell z )}{\ell z}.
  	\end{equation}  
  	
  	\subsubsection{Scalar Flux / Fluence}
  	Assuming a unit mean-free path $\ell = 1$ and substituting the transformed propagator $\zeta_3(z)$ into Equation~\ref{eq:phiz} and applying the inverse Fourier transform produces the well-known~\cite{bothe42,nuyens49,grosjean51,case53} expression for the scalar flux about a point source in an infinite medium
  	\begin{equation}\label{eq:casephi}
      \phipt(r) = \frac{\e^{-r}}{4 \pi r^2} + \frac{1}{2 \pi^2 r} \int_0^\infty \frac{\ssalbedo \, \text{arctan}^2\, z}{z-\ssalbedo \, \text{arctan} \, z} \, \sin (r \, z) \, dz.
    \end{equation}
    This oscillatory integral, though an exact solution, is not particularly well suited for numerical application.  Two equivalent, exact, and numerically-stable solutions are also known.  The first arises from a singular-eigenfunction analysis of the transport operator~\cite{case67,mccormick73} in plane-parallel infinite media (after applying the plane-to-point transformation~\cite{bell70}) and results in
      \begin{equation}\label{eq:phiptcase}
        \phipt(r) = \frac{1}{4 \pi r} \left[ \frac{\e^{-r /\nu_0}}{\nu_0 \, N_0^+} + \int_0^1 \frac{\e^{-r / \nu}}{\nu \, N_\nu} d\nu \right]
      \end{equation}
    where the quantities $N_0^+$ and $N_\nu$ are normalization integrals of the eigenfunctions of the transport operator in plane geometry~\cite{case67}, specifically
      \begin{align}
        N_0^{\pm} &= \pm \frac{\ssalbedo}{2} \nu_0^3 \left[ \frac{\ssalbedo}{\nu_0^2 - 1} - \frac{1  }{\nu_0^2} \right], \\  
        N_\nu &= \nu \left[ \lambda^2(\nu) + \frac{\pi^2 \ssalbedo^2}{4} \nu^2 \right], \\  
        \lambda(\nu) &= 1 - \ssalbedo \nu \arctan \nu
      \end{align}
    and the \emph{rigorous asymptotic diffusion length} $\nu_0$ is the positive real solution $\nu_0 > 1$ of the characteristic equation
     \begin{equation}\label{eq:characteristicExp3D}
      1 = \ssalbedo \nu_0 \arctan \frac{1}{\nu_0} = \frac{\ssalbedo \nu_0}{2} \log \frac{\nu_0 + 1}{\nu_0 - 1}.
    \end{equation}
    An equivalent result related to Equation~\ref{eq:phiptcase} by a change of variable $\nu=1/y$ was derived earlier by Davison in his 1943 seminal work~\cite{davison00} on the point source problem with isotropic scattering,
      \begin{equation}\label{eq:davisonphipt}
        \phipt(r) = \frac{1}{4 \pi r} \left[ \frac{\e^{-r/\nu_0}}{\nu_0 \, N_0^+} + \int_1^\infty \frac{ \e^{-r y}  }{\frac{\pi^2 \ssalbedo^2}{4 y^2}+\left( 1- \frac{\ssalbedo}{2y} \log \frac{y+1}{y-1}\right)^2} dy \right]
      \end{equation}
      where the spectrum of the transport operator was also discussed.

    \subsubsection{Relation to the spectrum of the transport operator}
    	Equations~\ref{eq:phiptcase} and~\ref{eq:davisonphipt} can be derived directly from the Fourier inversion of Equation~\ref{eq:phizunstable} using contour integration~\cite{case53}.  However, an alternative derivation is possible using the known spectrum of the transport operator in \emph{plane geometry}~\cite{case67}.  The plane-geometry spectrum appears directly in the solution for the spherically-symmetric point source problem---the scalar flux is the sum of a discrete term that dominates far from the point source and a transient term that dominates near the source.  An exact solution containing \emph{angular} quanties $L(r,\omega)$ in plane geometry is possible using the sum of a discrete angular eigenfunction and a continuous superposition of singular angular eigenfunctions---the eigenvalues of which ($\nu_0$ and $\nu$) are exactly the same as those seen in the solution for the scalar flux about a point source (Equation~\ref{eq:phiptcase}).  However, in spherical geometry, the angular eigenfunctions disappear~\cite{ganapol03}.

    	Our present interest in compact diffusion approximations requires only the discrete spectrum of the transport operator.  Given solutions of the form in Equation~\ref{eq:phiptcase} the \emph{rigorous asymptotic diffusion approximation} is formed by simply dropping the portion corresponding to the continuous, singular portion of the spectrum (the integral).  For the case of isotropic scattering, the single remaining discrete term \changed{is}
    	\begin{equation}\label{eq:rigdiff}
    		\frac{1}{4 \pi r} \frac{\e^{-r/\nu_0}}{\nu_0 \, N_0^+} 
    	\end{equation}
    	\changed{and this term} encompasses the entire rigorous asymptotic diffusion solution.  The number of discrete terms appearing in the point source Green's function is a function of the shape of the scattering kernel (phase function) and also (as we discuss later) on the dimension of the transport space.  For classical transport in three dimensions, when the scattering law is more general than linearly anisotropic scattering, more than one discrete eigenmode appears in the solution, with more modes appearing as the scattering kernel becomes more peaked~\cite{case53,davison57,grosjean63b,siewert99b}.

      The rigorous diffusion length $\nu_0$ in Equation~\ref{eq:rigdiff} corresponds to the largest discrete eigenvalue in the spectrum of the transport operator and can be found by inspection given the Fourier transformed solution---the rigorous diffusion length $\nu_0$ is the inverse magnitude of a purely imaginary root of the denominator of the transformed collision density (or scalar flux), a real, positive solution of
      \begin{equation}\label{eq:denom}
        1 - \ssalbedo \o{\zeta}_d(i / \nu_0) = 0.
      \end{equation}
      Thus, without knowing the eigenfunctions of the transport operator, we can discern the discrete eigenvalues by solving Equation~\ref{eq:denom}.  For the case of exponential scattering in 3D, equation~\ref{eq:denom} is exactly the famous characteristic equation~(\ref{eq:characteristicExp3D}).

	\subsubsection{Rigorous vs. $P_1$ diffusion approximations}\label{sec:rigorous}
    Equation~\ref{eq:rigdiff} is identified with diffusion because it is proportional to a point source \emph{diffusion mode}.  The family of point source diffusion modes with \emph{diffusion length} $\nu$ are the inverse transforms
      \begin{equation}
        \Finv \left\{ \frac{1}{1+(z \nu)^2} \right\} = (2 \pi )^{-d/2} r^{1-\frac{d}{2}} \nu^{-\frac{d}{2}-1}
   K_{\frac{d-2}{2}}\left(\frac{r}{\nu}\right),
      \end{equation}
      where $K$ is the modified Bessel function of the second kind.  In three dimensions, this yields
      \begin{equation}
        \mathcal{F}_3^{-1} \left\{ \frac{1}{1+(z \nu)^2} \right\} = \frac{e^{-r / \nu}}{4 \pi  r \nu^2}.
      \end{equation}
		The rigorous diffusion approximation arises from the contribution of the pole in a contour integral formed from the the Fourier inversion of $\o{\Cpt}(z)$ and is specifically~\cite{grosjean63b}
    \begin{equation}
      \phipt(r) \approx \frac{\partial(-\chi^2)}{\partial c} \mathcal{F}_d^{-1} \left\{ \frac{1}{\chi^2} \frac{1}{1+(z / \chi)^2} \right\}
    \end{equation}
    where $\chi = 1 / \nu_0$ is the inverse diffusion length, a positive real solution of
    \begin{equation}
        1 - c \o{\zeta}_d(i \chi) = 0.
    \end{equation}

    The rigorous asymptotic diffusion approximation becomes accurate far from the source and better approximates more of the total scalar densities as the single-scattering albedo $c$ approches $1$.  Rigorous diffusion differs significantly in both asymptote and magnitude from the classical ($P_1$) diffusion approximation
    \begin{equation}\label{eq:exp3dclassicaldiff}
      \phipt(r) \approx \frac{3 e^{-\sqrt{3-3 c} r}}{4 \pi  r},
    \end{equation}
    which can be found by forming an order $\{0,2\}$ \changed{Pad\'{e}} approximant about $z = 0$ of the transformed solution and inverting the resulting diffusion mode.  For classical transport theory in 3D, the $\{0,2\}$ \changed{Pad\'{e}} approximant of the scalar flux is
		\begin{equation}
			\frac{\o{X}(z)}{1-\ssalbedo \o{\zeta}_d(z)} = \frac{\arctan(z)}{z \left(1-\frac{c \arctan(z)}{z}\right)} \approx \frac{3}{3-3 c+z^2}
		\end{equation}
		whose inverse Fourier transform yields Equation~\ref{eq:exp3dclassicaldiff}.  Given the \emph{mean square free path}
    \begin{equation}
      \langle s^2 \rangle = \int_0^\infty s^2 p(s) ds
    \end{equation}
    the classical $P_1$ diffusion length has also been shown~\cite{larsen11} to be
    \begin{equation}
      \nu_0 = \sqrt{\frac{D}{1-c}}
    \end{equation}
    with the classical $P_1$ diffusion coefficient $D$ for collision density given by
    \begin{equation}\label{eq:classicalD}
      D = \frac{\langle s^2 \rangle}{2 d}.
    \end{equation}
    
    The classical diffusion coefficient for collision density $D$ in Equation~\ref{eq:classicalD} was derived in 3D~\cite{larsen11} and we conjecture that it holds in general dimension where the dependence on dimension $d$ is as indicated.  The complete diffusion approximation for the total collision density with arbitrary free path distribution and dimension is then
    \begin{equation}\label{eq:P1Conjecture}
      \Cpt(r) \approx \frac{1}{1-c} \mathcal{F}_d^{-1} \left\{ \frac{1}{1+(z \nu_0)^2} \right\}.
    \end{equation}
		
\begin{figure*}
	      \centering
	      \subfigure[Low absorption - comparison]{\includegraphics[width=.45\linewidth]{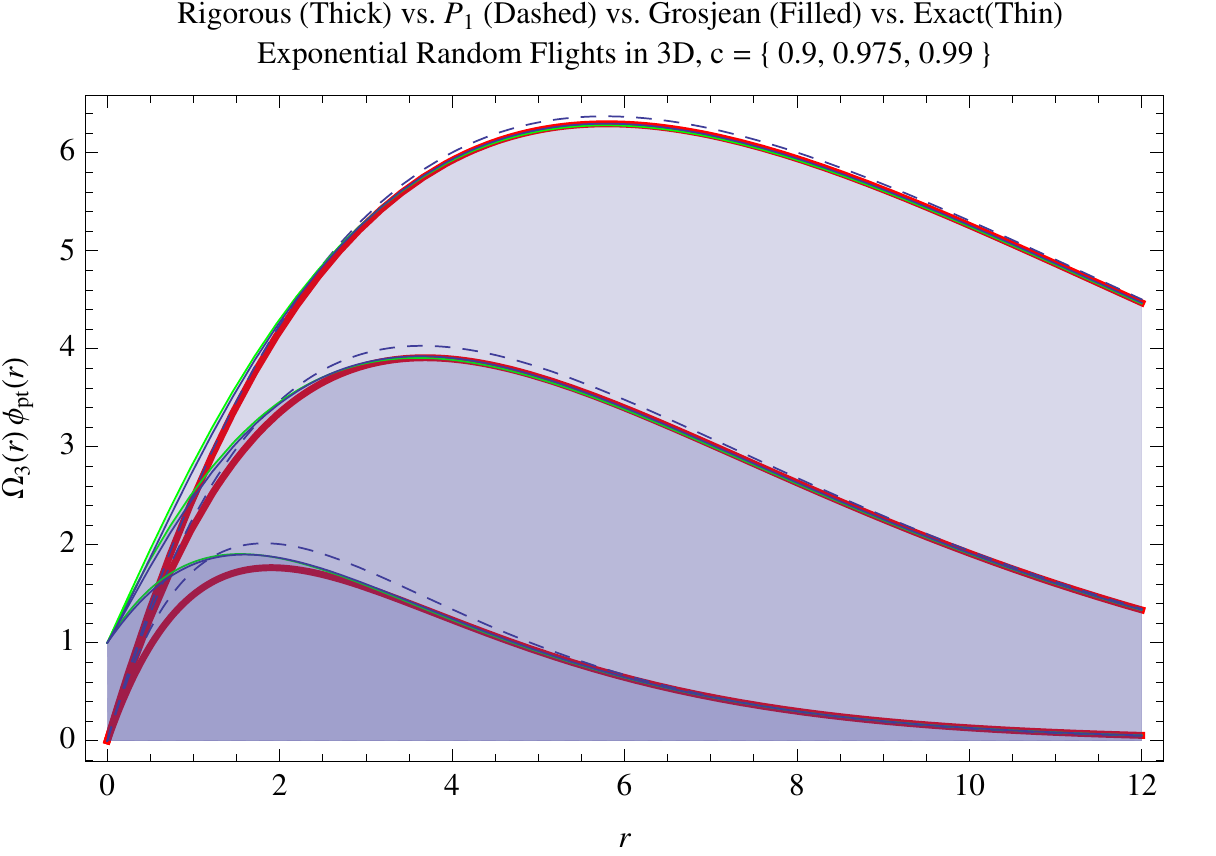}}
	      \subfigure[Low absorption - relative error]{\includegraphics[width=.48\linewidth]{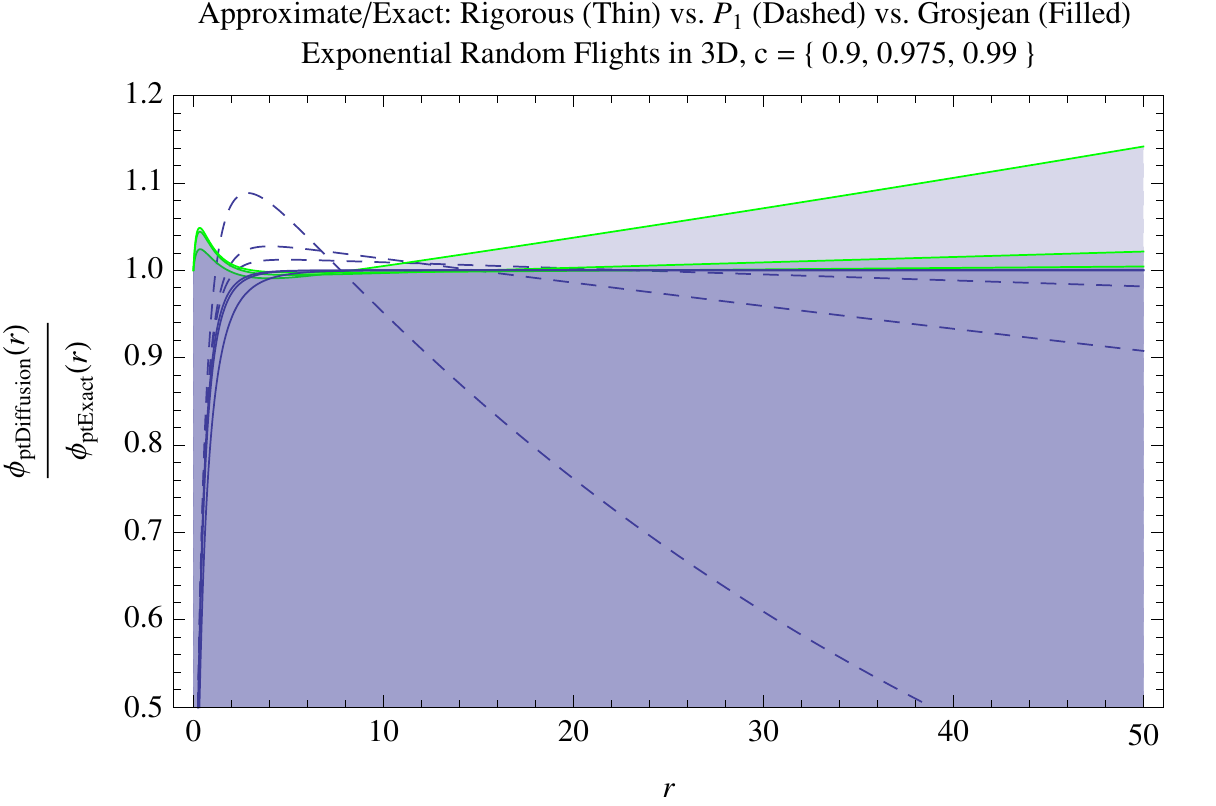}}
        \subfigure[High absorption - comparison]{\includegraphics[width=.45\linewidth]{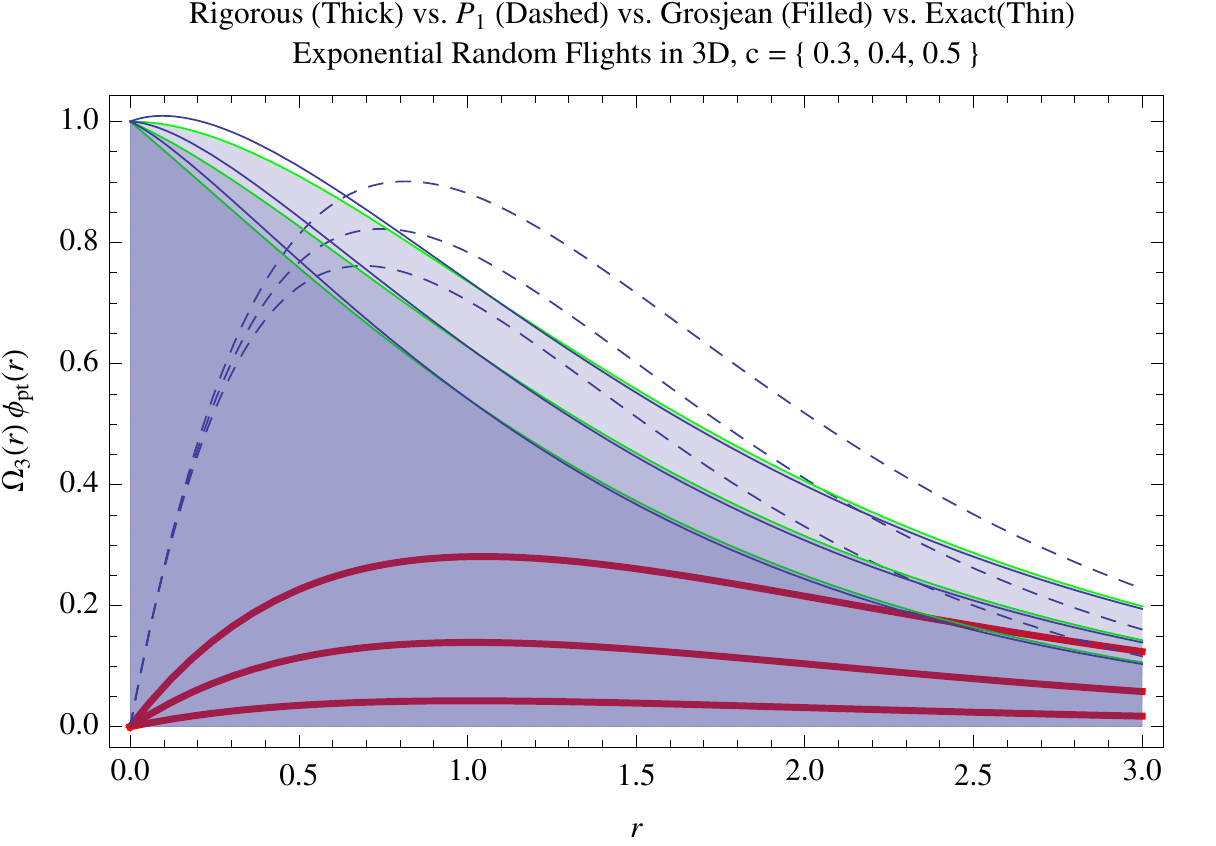}}
        \subfigure[High absorption - relative error]{\includegraphics[width=.48\linewidth]{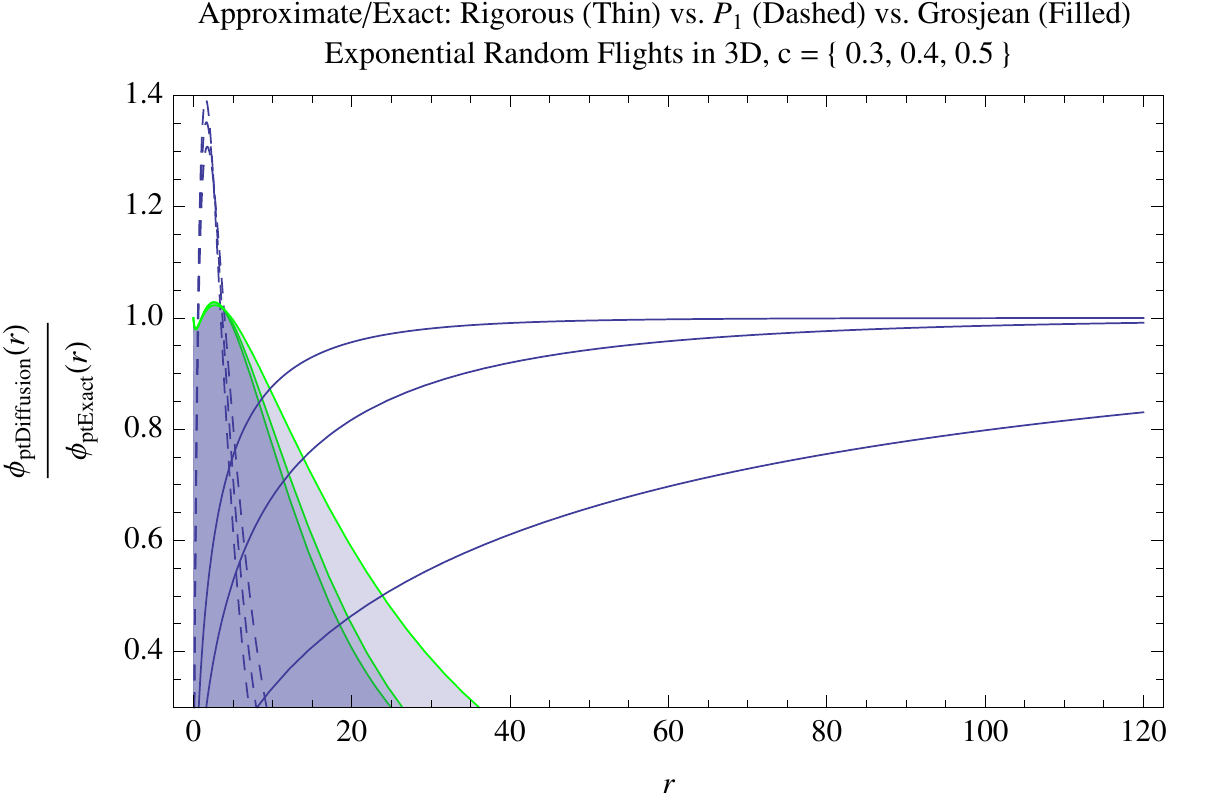}}
	      \caption{Comparison of the accuracy of classical ($P_1$), rigorous asymptotic and Grosjean-modified diffusion approximations for the scalar flux $\phipt(r)$ about an isotropic point source in an infinite 3D medium.  The rigorous and Grosjean forms of diffusion approximation that we generalize in this paper outperform classical $P_1$ diffusion near the source (Grosjean) or very far from the source where the density is low (rigorous).}
	      \label{fig-asymptClassicalCompare} 
	    \end{figure*}

    Figure~\ref{fig-asymptClassicalCompare} recalls the well-known improved performance of the rigorous vs. classical ($P_1$) diffusion for very low densities far from the point source as well as the overall breakdown of either approximation as the absorption level increases.  However, where $P_1$ diffusion has increasingly inaccurate asymptotics for high absorption levels, rigorous diffusion will eventually always become an accurate approximation, but this may require extremely large distances from the source (and extremely low densities).
  \subsubsection{Spatial Moments}\label{sec:moments}
    The spatial moments of a given distribution are useful for validating and deriving new approximations, as well as providing simple analytic benchmarks for testing Monte Carlo estimators.  Even-order moments of some radially symmetric function given by $f(r)$ are easily obtained from the derivatives of the spherical-Fourier-transformed distribution~\cite{zoia11e}
    \begin{equation}\label{eq:moments}
      \int_0^\infty r^m \Omega_d(r) f(r) dr = \frac{\sqrt{\pi } \Gamma \left(\frac{d+m}{2}\right)}{\Gamma
   \left(\frac{d}{2}\right) \Gamma \left(\frac{m+1}{2}\right)} \left[ \frac{\partial^m \o{f}(z)}{\partial (i \, z)^m} \right]_{z=0}
    \end{equation}
    where $\o{f}(z)$ is the transform $\o{f}(z) = \F \left\{ f(r) \right\} $ of $f(r)$.  Temporarily relaxing the assumption of a unit mfp $\ell = 1$, recall that the normalization of the free path distribution $p(s)$ implies $\o{\zeta_d}(0) = 1$~\cite{zoia11e}, and so the zeroth moment of the collision density is always
    \begin{equation}
      \int_0^\infty \Omega_d(r) \Cpt(r) dr = \frac{1}{1-\ssalbedo}
    \end{equation}
    regardless of free path distribution and dimension.  For the scalar flux / fluence in 3D with exponential free-paths, we have
    \begin{equation}
      \int_0^\infty \Omega_d(r) \phipt(r) dr = \frac{\ell}{1-\ssalbedo} = \frac{1}{\Sigma_a}
    \end{equation}
    where $\Sigma_a$ is the absorption coefficient.  

    For classical transport in three dimensions the even moments of the collision density have no known simple expression for arbitary $m$ but are found readily using Equation~\ref{eq:moments}.  We recall here the first few even moments of the $n$th collision density
    \begin{align}\label{eq:exp3dmomentsnth}
      &\int_0^\infty 4 \pi r^2 \Cpt(r|n) dr = c^{n-1} \\
      &\int_0^\infty 4 \pi r^4 \Cpt(r|n) dr = 2 c^{n-1} n \\
      &\int_0^\infty 4 \pi r^6 \Cpt(r|n) dr = \frac{4}{3} n (5 n+13) c^{n-1} \\
      &\int_0^\infty 4 \pi r^8 \Cpt(r|n) dr = \frac{8}{9} n \left(35 n^2+273 n+502\right) c^{n-1}
    \end{align}
    and the total collision density moments
    \begin{align}\label{eq:exp3dmoments}
      &\int_0^\infty 4 \pi r^2 \Cpt(r) dr = \frac{1}{1-c} \\
      &\int_0^\infty 4 \pi r^4 \Cpt(r) dr = \frac{2}{(c-1)^2} \\
      &\int_0^\infty 4 \pi r^6 \Cpt(r) dr = \frac{8 (4 c-9)}{3 (c-1)^3} \\
      &\int_0^\infty 4 \pi r^8 \Cpt(r) dr = \frac{16 \left(44 c^2-144 c+135\right)}{3 (c-1)^4}
    \end{align}
    which agree with previously reported expressions~\cite{grosjean51,case53,case67}.  For exponential free-path distributions the scalar flux / fluence moments are related to the collision density moments by a factor of the mean free path $\ell$.

	\subsubsection{Grosjean's uncollided-plus-diffusion approximation}\label{sec:grosjean}
    	Grosjean~\shortcite{grosjean56a} proposed seeking an approximation of the scalar flux about an isotropic point source in an infinite medium by beginning with the exact, uncollided flux explicitly and describing the remaining collided flux with a diffusion mode:
      \begin{equation}\label{eq:grosjeanansatz}
        \phipt(r) \approx \frac{\e^{-r}}{4 \pi r^2} + w_G \frac{ e^{-r / \nu_G}}{4 \pi r}
      \end{equation}
      where the magnitude $w_G$ of the diffusion mode and the diffusion length $\nu_G$ are determined by requiring that Equation~\ref{eq:grosjeanansatz} satisfy the first two even spatial moment relations (Equations~\ref{eq:exp3dmoments}).  Alternatively, the moment preservation is equivalent~\cite{grosjean63b} to inverting a $\{0,2\}$ \changed{Pad\'{e}} approximant about $z=0$ of the transformed, scattered-only scalar flux, which yields
      \begin{equation}
        \frac{\o{X}(z) \ssalbedo \o{\zeta}_d(z) }{1-\ssalbedo \o{\zeta}_d(z)} = \frac{c (\arctan z)^2}{z^2 \left(1-\frac{c \arctan z}{z}\right)} \approx \frac{3 c}{3 (1-c)+(2-c) z^2},
      \end{equation}
      which inverts into the diffusion mode
      \begin{equation}\label{eq:collidedgrosjean3dexp}
        \frac{3 c e^{-\sqrt{\frac{3}{c-2}+3} r}}{4 \pi  (2-c) r},
      \end{equation}
      which, when combined with the exact uncollided scalar flux, yields the complete approximation
      \begin{equation}\label{eq:grosjeanfinal3d}
        \phipt(r) \approx \frac{\e^{-r}}{4 \pi r^2} + \frac{3 c e^{-\sqrt{\frac{3}{c-2}+3} r}}{4 \pi  (2-c) r},
      \end{equation}
      which is Grosjean's original result~\cite{grosjean54}.

      Classical $P_1$ diffusion also exactly satisfies the first two even spatial moments of the collision distribution.  However, unlike in Grosjean's method, in $P_1$ diffusion, the sole diffusion mode must attempt to encompass the uncollided portion of the flux (for which the angular flux is singular, not diffuse).  By relaxing this requirement, the resulting approximation in Equation~\ref{eq:grosjeanfinal3d} is able to provide remarkably accurate results even for very large absorption levels and for regions near the source (Figure~\ref{fig-asymptClassicalCompare}).

      Like classical $P_1$ diffusion in 3D, we conjecture that Grosjean's modified diffusion length can be found without using Fourier transforms by using only the mean square free distance $\langle s^2 \rangle$,
      \begin{equation}\label{eq:grosjeanConjecture}
        \nu_G = \sqrt{\frac{\langle s^2 \rangle (2-c)}{2 \, d (1 - c)}},
      \end{equation}
      with the complete Grosjean diffusion approximation for the total collision density
      \begin{equation}
        \Cpt(r) \approx \frac{p(s)}{\Omega_d(r)} + \frac{c}{1-c} \mathcal{F}_d^{-1} \left\{ \frac{1}{1+(z \nu_G)^2} \right\}.
      \end{equation}
      We do not prove this conjecture here, but found it to hold for a large variety of derivations that follow, wherever we were able to compute the associated \changed{Pad\'{e}} approximants explicitly.

      We only discuss the infinite medium point source problem in this paper, but Grosjean further proposed extending this modified diffusion theory to non-infinite homogeneous problems~\cite{grosjean58a,grosjean58b} and it was later applied to the searchlight problem~\cite{deon11a,deon13a}.  Later in the paper we extend Grosjean's diffusion-based formalism to generalized Boltzmann linear transport in arbitrary dimensions.

    \subsubsection{Higher order \changed{Pad\'{e}}-approximant diffusion}
      Grosjean~\shortcite{grosjean63b} also proposed forming an improved approximation for the scalar flux about a point source by beginning with an uncollided plus double-diffusion ansatz where the rigorous asymptotic diffusion appears as one of the two diffusion modes and the second diffusion mode is found by having the total approximation satisfy the first three even moments of the exact solution.  We do not generalize this approach here, but note the potential for building a single approximation that has the benefits of both the rigorous asymptotic and Grosjean's original approximations.  

      We investigated building order $\{2,4\}$ approximants of the multiple-collision density.  Expanding the result into partial fractions yields two diffusion modes.  This can be continued to higher orders in theory, but factoring the polynomials becomes a road block.  However, MATHEMATICA can often invert higher order approximants when $c$ is fixed at a specific floating point value.  For example, multi-diffusion approximations for the collision density with $c = 0.9$ are,
      \begin{multline}
        \Cpt(r) \approx \frac{\e^{-r}}{4 \pi r^2} + \frac{0.180578 \e^{-0.525406 r}}{r}-\frac{0.0147796 \e^{-1.2429 r}}{r} \\
        \approx \frac{\e^{-r}}{4 \pi r^2} -\frac{0.00921106 e^{-1.79529 r}}{r}-\frac{0.00817961 e^{-1.08754 r}}{r} \\+\frac{0.180651 e^{-0.525429 r}}{r} \\
        \approx \frac{\e^{-r}}{4 \pi r^2} -\frac{0.00647263 e^{-2.39812 r}}{r}-\frac{0.00695338 e^{-1.32319 r}}{r} \\-\frac{0.00511902 e^{-1.04597 r}}{r}+\frac{0.180651 e^{-0.52543 r}}{r}
      \end{multline}
      but we found these provided no measurable accuracy improvement.

    \subsubsection{Deriving Approximate $n$th-collided solutions}

\begin{figure*}
        \centering
        \includegraphics[width=.98\linewidth]{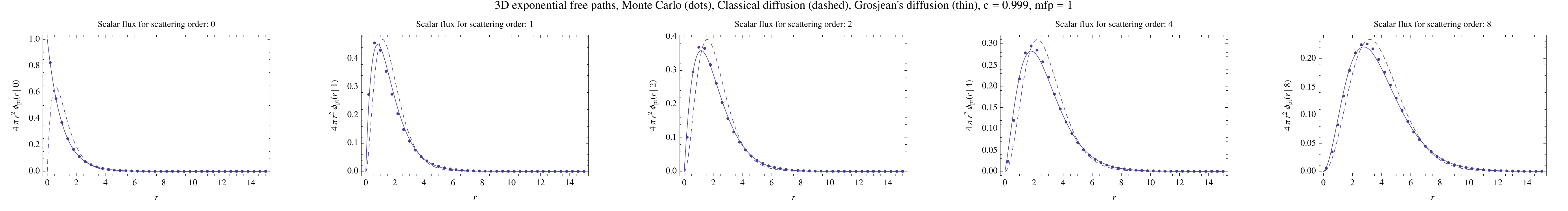}
        \caption{Exact and approximate $n$th-collided scalar flux distributions about an isotropic point source in 3D with exponential random flights.  The approximate solutions result from forming a Taylor series expansion about $c = 0$ of the classical and Grosjean diffusion approximations and provide compact analytic forms (for low $n$).}
        \label{fig:perorderdiffcompare} 
      \end{figure*}

      It has been argued~\cite{guth60} that random-flight theory delivers more information about the scattering process than linear transport theory by explicitly providing the collision-order quantities $\Cpt(r|n)$ and $\phi(r|n)$.  However, any analytic, infinitely-differentiable result, exact or approximate, contains implicitly, via its dependence on the single-scattering albedo $\ssalbedo$, a full set of collision-order solutions.  \emph{Any analytic, infinitely-differentiable multiple-scattering quantity $f(c)$ can be used to form an $n$th-collided expansion by forming the Taylor series expansion of $f(c)$ about $c = 0$}.  For example, the $n$th-collided scalar fluxes contained within the classical diffusion approximation (Equation~\ref{eq:exp3dclassicaldiff}) are
      \begin{equation}
        \phipt(r) \approx \frac{3 e^{-\sqrt{3} r}}{4 \pi  r}  +  c \, \frac{3 \sqrt{3} e^{-\sqrt{3} r}}{8 \pi } +  c^2 \, \frac{3 e^{-\sqrt{3} r} \left(3 r+\sqrt{3}\right)}{32 \pi }  + ... .
      \end{equation}
      Similarly, the $n$th-collided scalar fluxes formed from Grosjean's diffusion-based approximation are
      \begin{equation}
        \phipt(r) \approx \frac{e^{-r}}{4 \pi r^2} +c \, \frac{3 e^{-\sqrt{\frac{3}{2}} r}}{8 \pi  r}+ c^2 \, \frac{3 e^{-\sqrt{\frac{3}{2}} r} \left(\sqrt{6} r+4\right)}{64 \pi  r}+ ... .
      \end{equation}
      We can compare the accuracy of these results to the exact solutions, which are only known in integral form.  The Fourier domain analysis yields the $n$th collision densities~\cite{placzek43}
      \begin{align}
        \Cpt(r|n>0) &= \Finv \{ \left(\ssalbedo \frac{ \arctan z}{z} \right)^n \} = \\
        & \int_0^\infty \frac{z \sin (r z) \left(\frac{c \arctan(z)}{z}\right)^n}{2 \pi ^2 r} dz.
      \end{align}
      An alternative integral form of this exact solution is~\cite{grosjean51}
      {\small
      \begin{multline*}\label{eq:phinGrosjean}
        \Cpt(r|n > 0) = \\ \frac{\ssalbedo^n}{2^{n+2} \pi^2 r \, i} \int_1^\infty \frac{ e^{-r z}}{z^{n-1}} \left(\left(\log \left(\frac{z+1}{z-1}\right)+i \pi
  \right)^{n}-\left(\log \left(\frac{z+1}{z-1}\right)-i \pi
  \right)^{n}\right) dz.
      \end{multline*}}
      Figure~\ref{fig:perorderdiffcompare} compares the classical- and Grosjean-diffusion-  $n$th-collided scalar fluxes to exact Monte Carlo reference solutions.  The Grosjean approximation produces the exact uncollided term explicitly, and the $n$-th collided terms are more accurate than those extracted from the classical $P_1$ diffusion approximation.  For large $n$ it is common~\cite{grosjean51,zoia11e} to extrapolate Gaussian approximations for $n$th-collided distributions from a Taylor expansion of  transformed $n$th collided distributions.
\section{Exponential Random Flights ($d \ne 3$)}\label{sec:exponential}

  We continue by considering classical transport theory (radiative transfer) with exponential free paths in dimensions other than 3.  For closely-related time-resolved solutions to these problems, we refer the reader to several related papers.  Exact time-resolved solutions are known in 2D~\cite{stadje87,stadje89,paasschens97} and 4D~\cite{paasschens97,orsingher07}.  Numerically integrating these time-resolved solutions over all times $t > 0$ provides additional formulations of the steady-state results.

  \subsection{Exponential---1D Rod Model case ($d$ = 1)}
    \subsubsection{Propagators}
      For exponential random flights with mean $\ell = 1$ in a 1D rod the transformed propagator and transformed stretched extinction are both
      \begin{equation}\label{eq:exprodprop}
        \o{\zeta}_1(z) = \o{X}(z) = \frac{1}{1+z^2}.
      \end{equation}

    \subsubsection{Collision Densities (Exponential $d = 1$)}
      The transformed $n$th-collision densities can be inverted to produce the exact analytic solutions
      \begin{equation}
        \Cpt(r|n) = \frac{2^{\frac{1}{2}-n} \sqrt{r} (c r)^{n-1} K_{\frac{1}{2}-n}(r)}{\sqrt{\pi } \Gamma (n)}.
      \end{equation}
      and similarly for the total collision density
      \begin{equation}
        \Cpt(r) = \frac{1}{2 \sqrt{1-\ssalbedo} } \e^{-r \sqrt{1-\ssalbedo}}.
      \end{equation}
      These $n$th-collision densities can be analytically integrated to produce the spatial moments of any order $m$
      \begin{equation}
        \int_{0}^{\infty} r^m \Omega_d(r) \Cpt(r|n) dr = \frac{2^m c^{n-1} \Gamma \left(\frac{m+1}{2}\right) \Gamma \left(\frac{m}{2}+n\right)}{\sqrt{\pi } \Gamma (n)}
      \end{equation}
      and, summing over all collision orders $n \ge 1$,
      \begin{equation}
        \int_{0}^{\infty} r^m \Omega_d(r) \Cpt(r) dr = (1-c)^{-\frac{m}{2}-1} \Gamma (m+1).
      \end{equation}

    \subsubsection{Scalar Flux / Fluence (Exponential $d = 1$)}
      The transformed $n$th-collided scalar fluxes can be inverted to produce the exact analytic solutions
      \begin{equation}
        \phipt(r|n) = \frac{2^{-n-\frac{1}{2}} \sqrt{r} (c r)^n K_{-n-\frac{1}{2}}(r)}{\sqrt{\pi } \Gamma (n+1)}.
      \end{equation}
      The scalar flux / fluence is also known exactly~\cite{wing62,zoia11e}
      \begin{equation}\label{rodexpphi}
        \phipt(r) = \frac{1}{2 \sqrt{1-\ssalbedo} } \e^{-r \sqrt{1-\ssalbedo}}.
      \end{equation}
      Similar to the case for the collision densities, for the scalar flux / fluence, the $n$th-collided moments are
      \begin{equation}
        \int_{0}^{\infty} r^m \Omega_d(r) \phipt(r|n) dr = \frac{2^m c^n \Gamma \left(\frac{m+1}{2}\right) \Gamma \left(\frac{m}{2}+n+1\right)}{\sqrt{\pi } \Gamma (n+1)}
      \end{equation}
      and for the total scalar flux / fluence,
      \begin{equation}
        \int_{0}^{\infty} r^m \Omega_d(r) \phipt(r) dr = (1-c)^{-\frac{m}{2}-1} \Gamma (m+1)
      \end{equation}

    \subsubsection{Diffusion Approximations (Exponential $d = 1$)}
      Diffusion with diffusion length $1 / \sqrt{1-\ssalbedo}$ is an exact solution for exponential random flights in a one-dimensional rod.  Thus, equation~\ref{rodexpphi} is both the classical $P_1$ and rigorous asymptotic diffusion solution.

  \subsection{Exponential---Flatland ($d = 2$)}
    \subsubsection{Propagators}
      For exponential random flights with mean $\ell = 1$ in flatland the transformed propagator and transformed stretched extinction are both
      \begin{equation}
        \o{\zeta}_2(z) = \o{X}(z) = \frac{1}{\sqrt{z^2+1}}.
      \end{equation}

    \subsubsection{Collision Densities (Exponential $d = 2$)}
      The transformed $n$th-collision densities can be inverted to produce the exact analytic solutions~\cite{stadje87,zoia11e}
      \begin{equation}
        \Cpt(r|n) = \frac{2^{-\frac{n}{2}-1} n c^{n-1} r^{\frac{n}{2}-1} K_{1-\frac{n}{2}}(r)}{\pi \Gamma \left(\frac{n}{2}+1\right)}.
      \end{equation}
      The case of $n = 3$ exhibits the remarkable property that in flatland a three-step exponential random flight of total length $t$ produces a uniform distribution of collisions inside the disc of radius $t$~\cite{franceschetti07}.  The time-resolved solution is thus proportional to $r$, and the steady-state solution is a simple exponential,
      \begin{equation}
        \Cpt(r|3) = \frac{c^2 e^{-r}}{2 \pi }.
      \end{equation}
      We were unable to analytically invert the total collision density
      \begin{equation}
        \Cpt(r) = \frac{\e^{- r}}{2 \pi r} + \frac{1}{2 \pi } \int_0^{\infty } \frac{\ssalbedo z J_0(r z)}{1 + z^2 - c \sqrt{1+z^2}} \, dz.
      \end{equation}
      The $n$th-collision densities can be analytically integrated to produce the spatial moments of any order $m$
      \begin{equation}
        \int_{0}^{\infty} r^m \Omega_d(r) \Cpt(r|n) dr = \frac{2^m c^{n-1} \Gamma \left(\frac{m}{2}+1\right) \Gamma
   \left(\frac{m+n}{2}\right)}{\Gamma \left(\frac{n}{2}\right)}
      \end{equation}
      and, summing over all collision orders $n \ge 1$, we find for the total collision density moments,
      \begin{multline}
        \int_{0}^{\infty} r^m \Omega_d(r) \Cpt(r) dr =\\ \left(1-c^2\right)^{-\frac{m}{2}-1} \left(m! \,
   _2F_1\left(-\frac{1}{2},-\frac{m}{2};\frac{1}{2};c^2\right)+c 2^m \Gamma
   \left(\frac{m}{2}+1\right)^2\right)
      \end{multline}
      where $_2F_1$ is a hypergeometric function.  

    \subsubsection{Scalar Flux / Fluence (Exponential $d = 2$)}
      Likewise, the transformed $n$th-collided scalar fluxes can be inverted to produce the exact analytic solutions
      \begin{equation}
        \phipt(r|n) = \frac{2^{-\frac{n}{2}-\frac{1}{2}} c^n r^{\frac{n-1}{2}} K_{\frac{1-n}{2}}(r)}{\pi  \Gamma \left(\frac{n+1}{2}\right)}.
      \end{equation}
      but an exact analytic inversion is not known for
      \begin{equation}
        \phipt(r) = \frac{\e^{- r}}{2 \pi r} + \frac{1}{2 \pi } \int_0^{\infty } \frac{\ssalbedo z J_0(r z)}{1 + z^2 - c \sqrt{1+z^2}} \, dz.
      \end{equation}
      Other known exact forms for the flatland scalar flux include
      \begin{equation}
        \phipt(r) = \frac{1}{2\pi} \int_0^\infty \frac{k J_0(r k)}{\sqrt{1+k^2}-c} dk
      \end{equation}
      and~\cite{zoia11e}
      \begin{equation}\label{eq:flatlandphizoia}
        \phipt(r) = \ssalbedo \frac{K_0(r \sqrt{1-\ssalbedo^2})}{\pi} + \frac{1}{2\pi} \int_0^\infty \frac{k J_0(r k)}{\sqrt{1+k^2}+c} dk
      \end{equation}
      and~\cite{liemert11c}
      \begin{equation}
        \phipt(r) = \frac{e^{-r}}{2 \pi  r} +\frac{c
   K_0\left(\sqrt{1-c^2} r\right)}{2 \pi }+ \frac{\sum _{n=1}^{\infty } \frac{2^{n+\frac{1}{2}} \sqrt{\frac{1}{r}} n! c^{2
   n} r^n K_{n-\frac{1}{2}}(r)}{\sqrt{\pi } (2 n)!}}{2 \pi }.
      \end{equation}

    \subsubsection{Diffusion Approximations (Exponential $d = 2$)}
      In flatland with exponential free-paths the characteristic equation (Equation~\ref{eq:denom}) is
      \begin{equation}
        1-\frac{c}{\sqrt{1-\frac{1}{\text{v0}^2}}}=0
      \end{equation}
      yielding the single discrete eigenvalue of the transport operator $\nu_0 = 1 / \sqrt{1-c^2}$.  The rigorous asymptotic diffusion approximation with this diffusion length 
      \begin{equation}\label{eq:flatlandexprig}
        \phipt(r) \approx \ssalbedo \frac{K_0(r \sqrt{1-\ssalbedo^2})}{\pi}
      \end{equation}
      appears in the expression derived recently by several authors (Equation~\ref{eq:flatlandphizoia}).  We note that a Taylor series expansion of Equation~\ref{eq:flatlandexprig} about $c = 0$ contains only odd powers of $c$, so, in some sense, the rigorous diffusion approximation in Flatland with exponential free paths represents only odd-ordered scattering events.

\begin{figure*}
        \centering
        \subfigure[Low absorption - comparison]{\includegraphics[width=.45\linewidth]{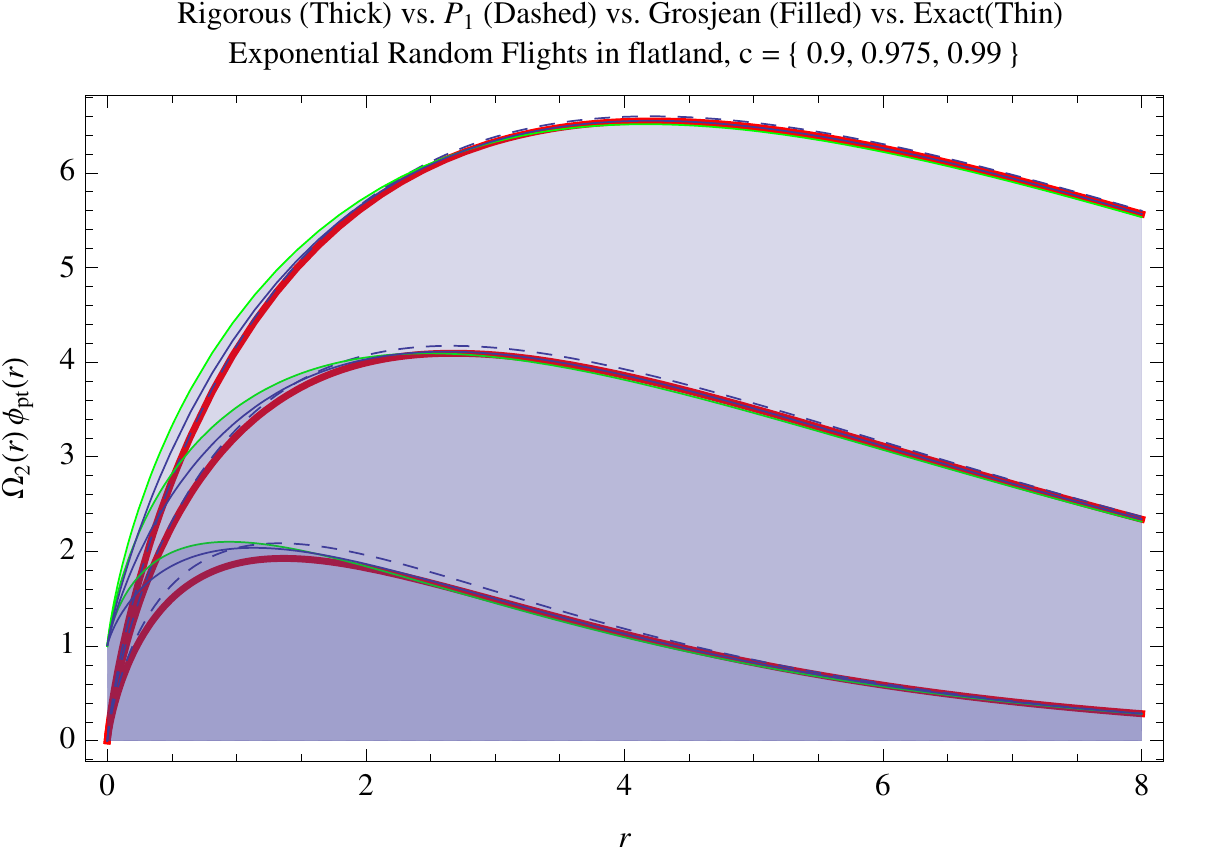}}
        \subfigure[Low absorption - relative error]{\includegraphics[width=.48\linewidth]{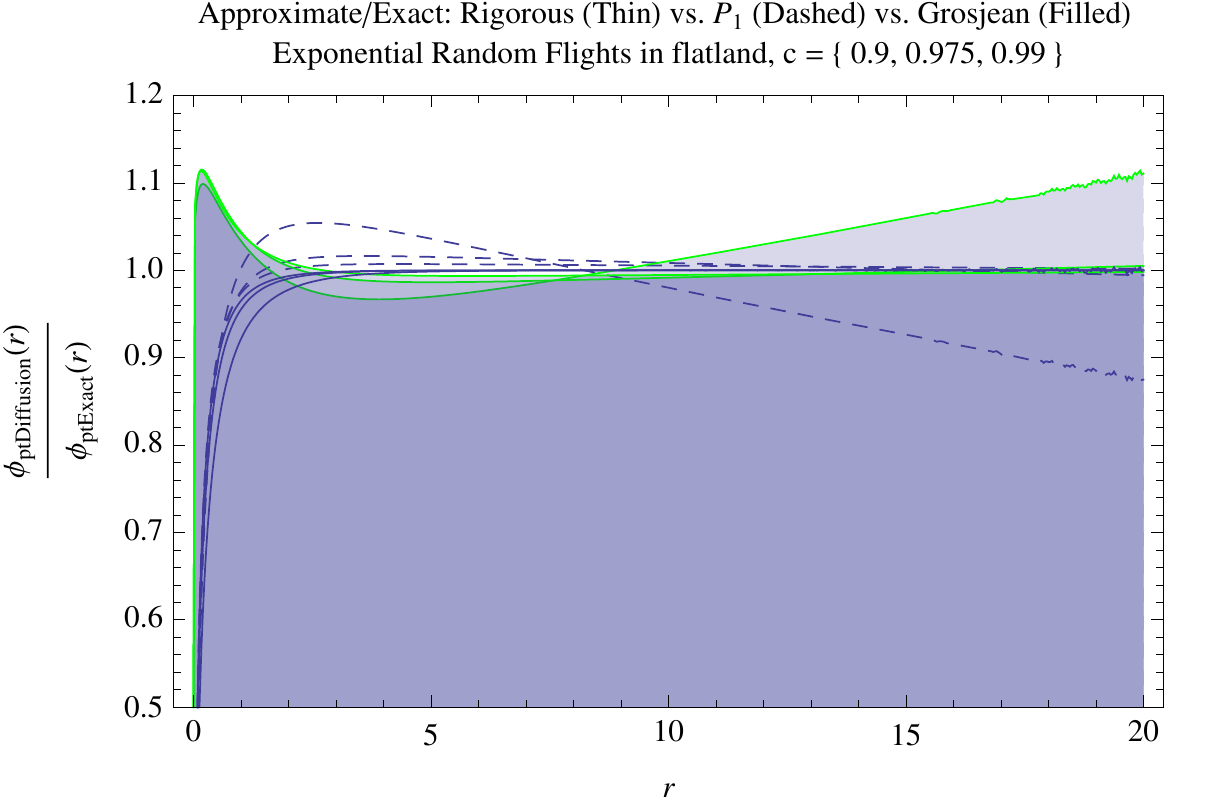}}
        \subfigure[High absorption - comparison]{\includegraphics[width=.45\linewidth]{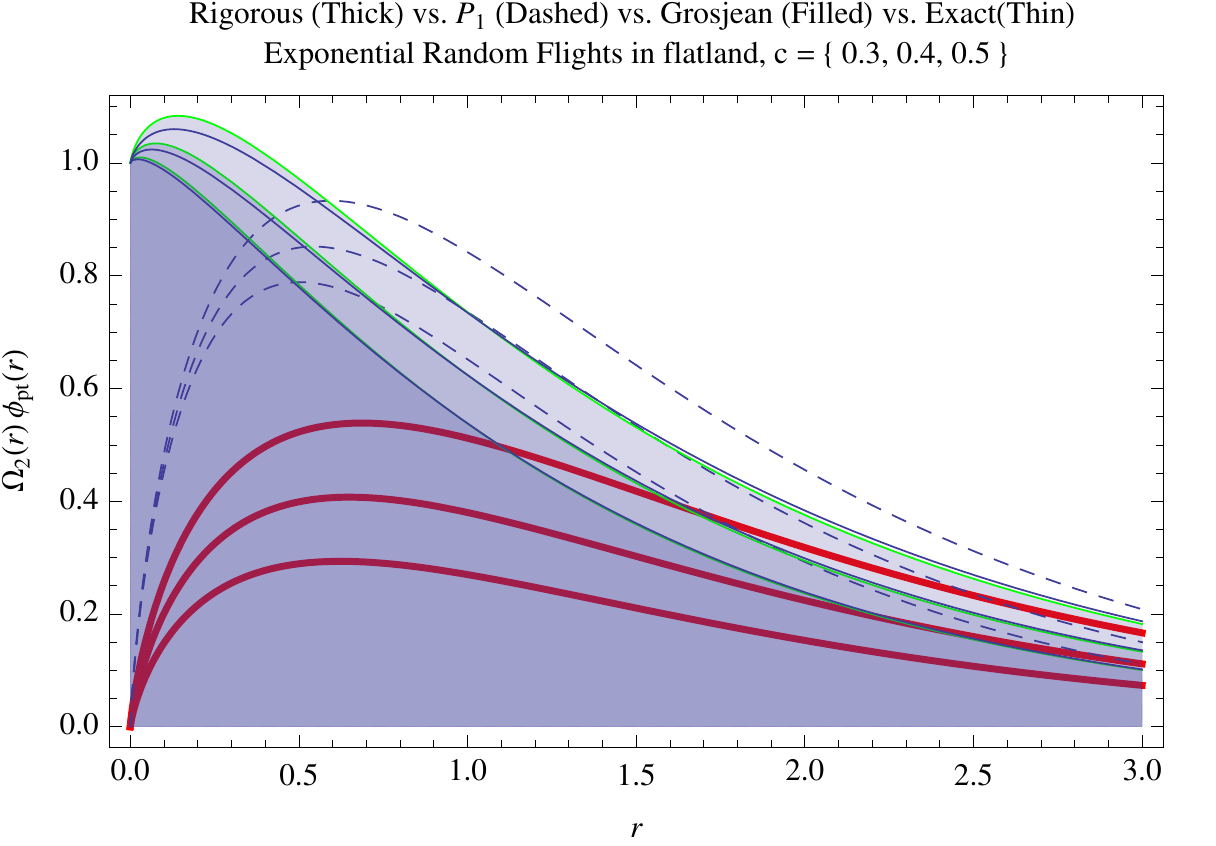}}
        \subfigure[High absorption - relative error]{\includegraphics[width=.48\linewidth]{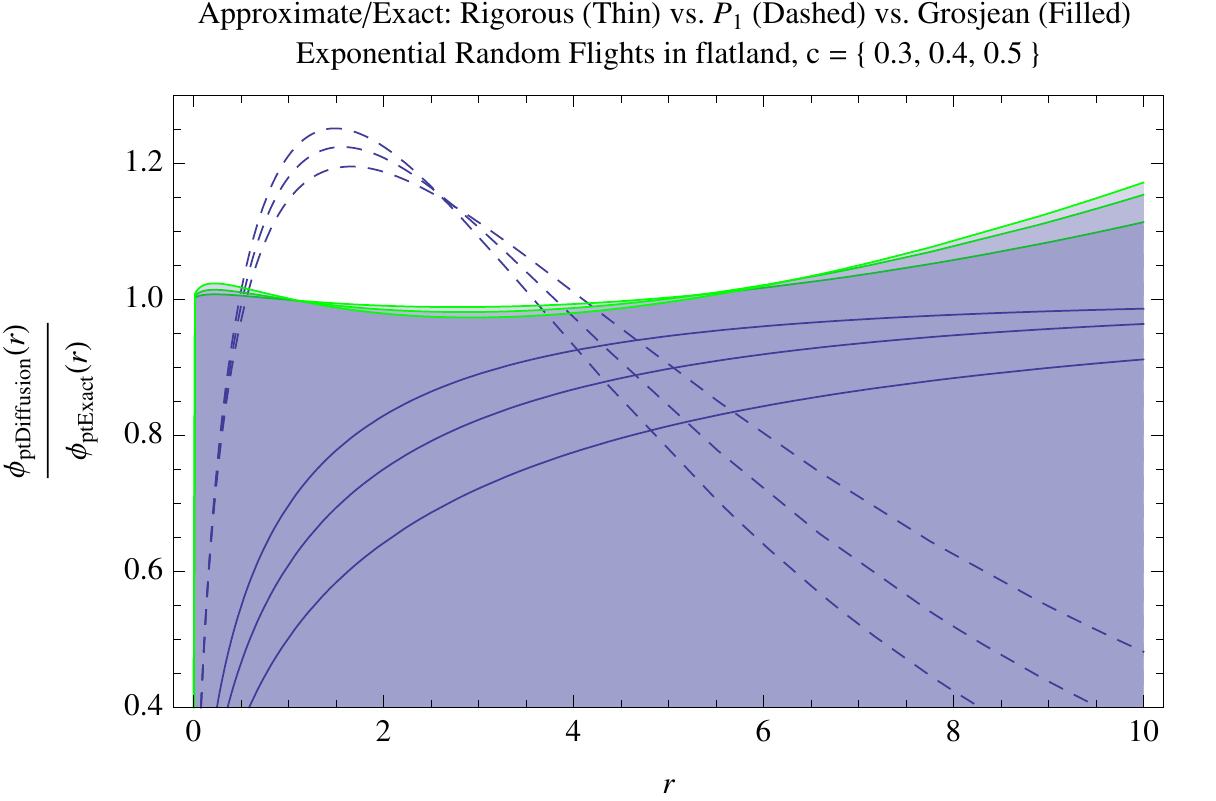}}
        \caption{Comparison of the accuracy of classical ($P_1$), rigorous asymptotic and Grosjean modified diffusion approximations for the scalar flux $\phipt(r)$ about an isotropic point source in an infinite medium in flatland.}
        \label{fig-expdiffusionCompareFlatland} 
      \end{figure*}

      The \changed{Pad\'{e}} approximant of the total transformed scalar flux is
      \begin{equation}
        \frac{c}{-c \sqrt{z^2+1}+z^2+1} \approx \frac{2}{-2 c+z^2+2},
      \end{equation}
      whose inversion produces the classical $P_1$ diffusion approximation in flatland
      \begin{equation}
        \phipt(r) \approx \frac{K_0\left(\sqrt{2-2 c} r\right)}{\pi }.
      \end{equation}
      The \changed{Pad\'{e}} approximant of the transformed collided scalar flux
      \begin{equation}
        \frac{c}{-c \sqrt{z^2+1}+z^2+1} \approx \frac{2 c}{(2-c) z^2+2 (1-c)}
      \end{equation}
      leads to the Grosjean modified-diffusion approximation in flatland
      \begin{equation}
        \phipt(r) \approx \frac{\e^{-r}}{2 \pi r} + \frac{c K_0\left(\sqrt{2+\frac{2}{c-2}} r\right)}{\pi  (2-c)}.
      \end{equation}
      Figure~\ref{fig-expdiffusionCompareFlatland} compares the various diffusion approximations for exponential random flights in flatland.  Similar to 3D, we see the Grosjean and rigorous diffusion approximations outperforming $P_1$.

  \subsection{Exponential Random flights in 4D ($d = 4$)}
    \subsubsection{Propagators}
      For exponential random flights with mean $\ell = 1$ in 4D the transformed propagator and transformed stretched extinction are both
      \begin{equation}
        \o{\zeta}_4(z) = \o{X}(z) = \frac{2 \left(\sqrt{z^2+1}-1\right)}{z^2}.
      \end{equation}

    \subsubsection{Collision Densities (Exponential $d = 4$)}
      In 4D with exponential free-paths the $n$th and total collision densities have no known analytic form but can be numerically inverted
      \begin{equation}
        \Cpt(r|n) = \frac{1}{4 \pi ^2 r} \int_0^\infty 2^n z^2 c^{n-1}
   \left(\frac{\sqrt{z^2+1}-1}{z^2}\right)^n J_1(r z) dz.
      \end{equation}
      The collision density for double collision, however, can be inverted
      \begin{equation}
        \Cpt(r|2) = -\frac{c \left(r^2 \text{Ei}(-r)+e^{-r} (r-1)\right)}{\pi ^2 r^2}
      \end{equation}
      where $Ei$ is the exponential integral. We were unable to analytically invert the total collision density
      \begin{equation}
        \Cpt(r) = \frac{\e^{- r}}{2 \pi^2 r^3} + \frac{1}{4 \pi^2 r} \int_0^{\infty } \frac{4 c \left(\sqrt{z^2+1}-1\right)^2 J_1(r z)}{z^2-2 c
   \left(\sqrt{z^2+1}-1\right)} \, dz.
      \end{equation}
      The first few even collision density moments are
      \begin{align}
        &\int_0^\infty \Omega_4(r)  \Cpt(r|n) dr = c^{n-1} \\
        &\int_0^\infty \Omega_4(r) r^2 \Cpt(r|n) dr = 2 c^{n-1} n \\
        &\int_0^\infty \Omega_4(r) r^4 \Cpt(r|n) dr = 6 n (n+3) c^{n-1} \\
        &\int_0^\infty \Omega_4(r) r^6 \Cpt(r|n) dr = 24 n \left(n^2+9 n+20\right) c^{n-1}
      \end{align}
      and the total collision density moments
      \begin{align}
        &\int_0^\infty \Omega_4(r) \Cpt(r) dr = \frac{1}{1-c} \\
        &\int_0^\infty \Omega_4(r) r^2 \Cpt(r) dr = \frac{2}{(c-1)^2} \\
        &\int_0^\infty \Omega_4(r) r^4 \Cpt(r) dr = \frac{12 (c-2)}{(c-1)^3} \\
        &\int_0^\infty \Omega_4(r) r^6 \Cpt(r) dr = \frac{144 \left(2 c^2-6 c+5\right)}{(c-1)^4}
      \end{align}

    \subsubsection{Scalar Flux / Fluence (Exponential $d = 4$)}
      Likewise, the transformed $n$th scattered scalar fluxes can be numerically inverted
      \begin{equation}
        \phipt(r|n) = \frac{1}{4 \pi ^2 r} \int_0^\infty 2^{n+1} z^2
   \left(\frac{c \sqrt{z^2+1}-1}{z^2}\right)^{n+1} J_1(r z) dz,
      \end{equation}
      with known exact solution for singly-scattered scalar flux
      \begin{equation}
        \phipt(r|1) = -\frac{c \left(r^2 \text{Ei}(-r)+e^{-r} (r-1)\right)}{\pi ^2 r^2}.
      \end{equation}
      We were unable to analytically invert the total scalar flux / fluence
      \begin{equation}
        \phipt(r) = \frac{\e^{- r}}{2 \pi^2 r^3} + \frac{1}{4 \pi^2 r} \int_0^{\infty } \frac{4 c \left(\sqrt{z^2+1}-1\right)^2 J_1(r z)}{z^2-2 c
   \left(\sqrt{z^2+1}-1\right)} \, dz.
      \end{equation}

      \paragraph{Exponential Random flights in 4D with $c = 1/2$}
        For the case of $c = 1/2$ the transformed scalar flux is invertible producing
        \begin{equation}\label{eq:4dexpexact}
          \phipt(r) = \frac{\e^{-r}}{2 \pi ^2 r^3} + \Finv \left\{ \frac{2-\frac{2}{\sqrt{z^2+1}}}{z^2} \right\} = \frac{\e^{-r}(1+r)}{2 \pi ^2 r^3}.
        \end{equation}

    \subsubsection{Diffusion Approximations (Exponential $d = 4$)}
      In 4D with exponential free-paths the characteristic equation (Equation~\ref{eq:denom}) is
      \begin{equation}
        2 c \left(\sqrt{1-\frac{1}{\text{v0}^2}}-1\right) \text{v0}^2+1=0
      \end{equation}
      yielding the single discrete eigenvalue of the transport operator, 
      \begin{equation}\label{eq:difflengthrig4d}
        \nu_0 = 1 / ( 2 \sqrt{c-c^2} ).
      \end{equation}
      The rigorous diffusion approximation in 4D is then
      \begin{equation}
        \phipt(r) \approx \frac{2 \sqrt{(1-c) c} (2 c-1) K_1\left(2 \sqrt{(1-c) c} r\right)}{\pi ^2 r}.
      \end{equation}
      For the first time, we see that in 4D the rigorous diffusion approximation is only valid for $c > 0.5$.  When the single-scattering albedo is exactly $c = 0.5$, the rigorous diffusion length is $\nu_0 = 1$ and the magnitude of the rigorous diffusion approximation goes to zero (where the exact solution is then solvable---see Equation~\ref{eq:4dexpexact}).  For $c < 0.5$ the rigorous diffusion length given by Equation~\ref{eq:difflengthrig4d} increases again, and the approximation breaks down completely.

\begin{figure*}
        \centering
        \subfigure[Low absorption - comparison]{\includegraphics[width=.45\linewidth]{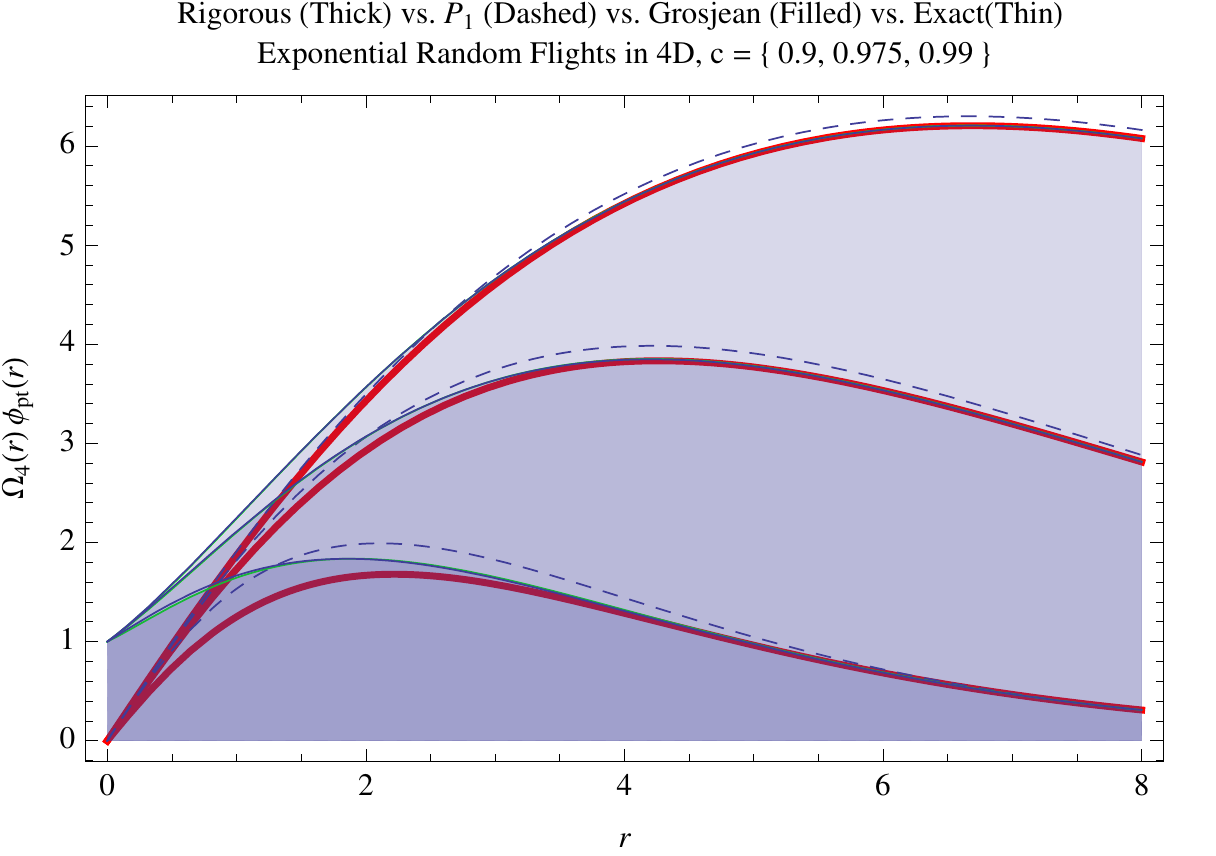}}
        \subfigure[Low absorption - relative error]{\includegraphics[width=.48\linewidth]{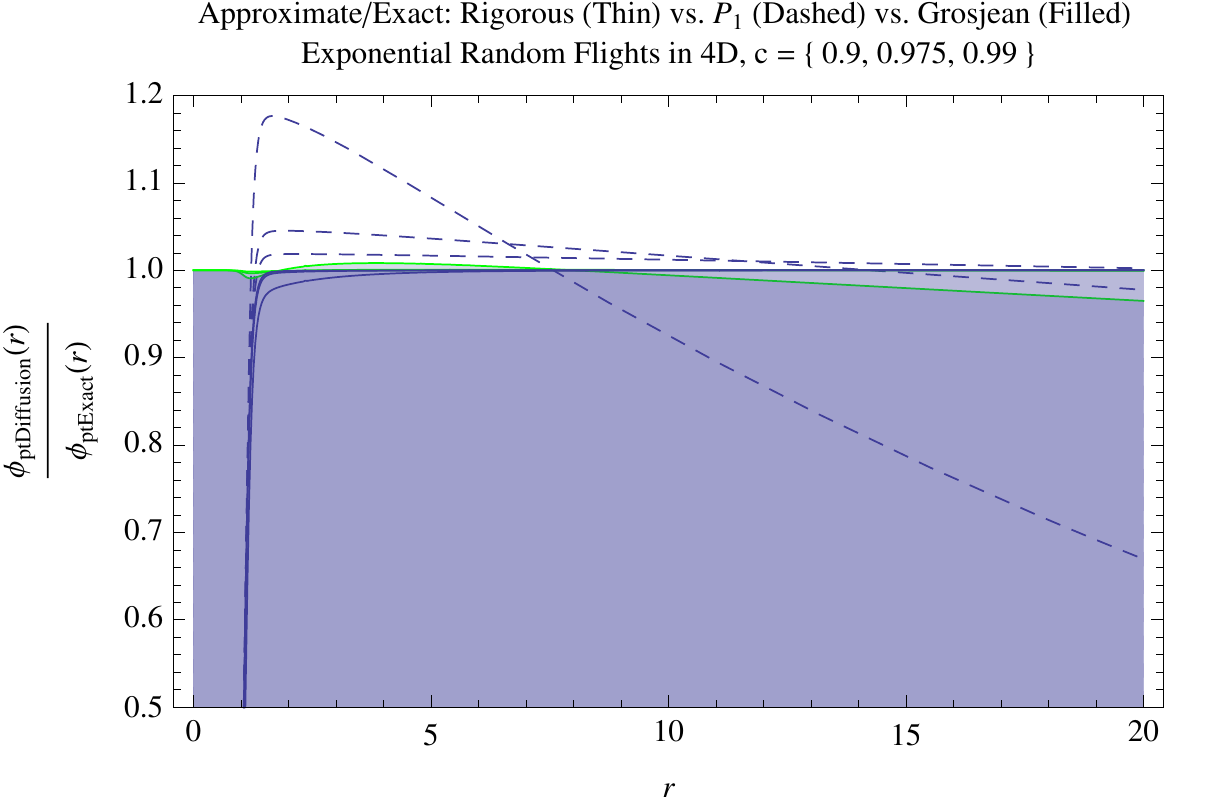}}
        \subfigure[High absorption - comparison]{\includegraphics[width=.45\linewidth]{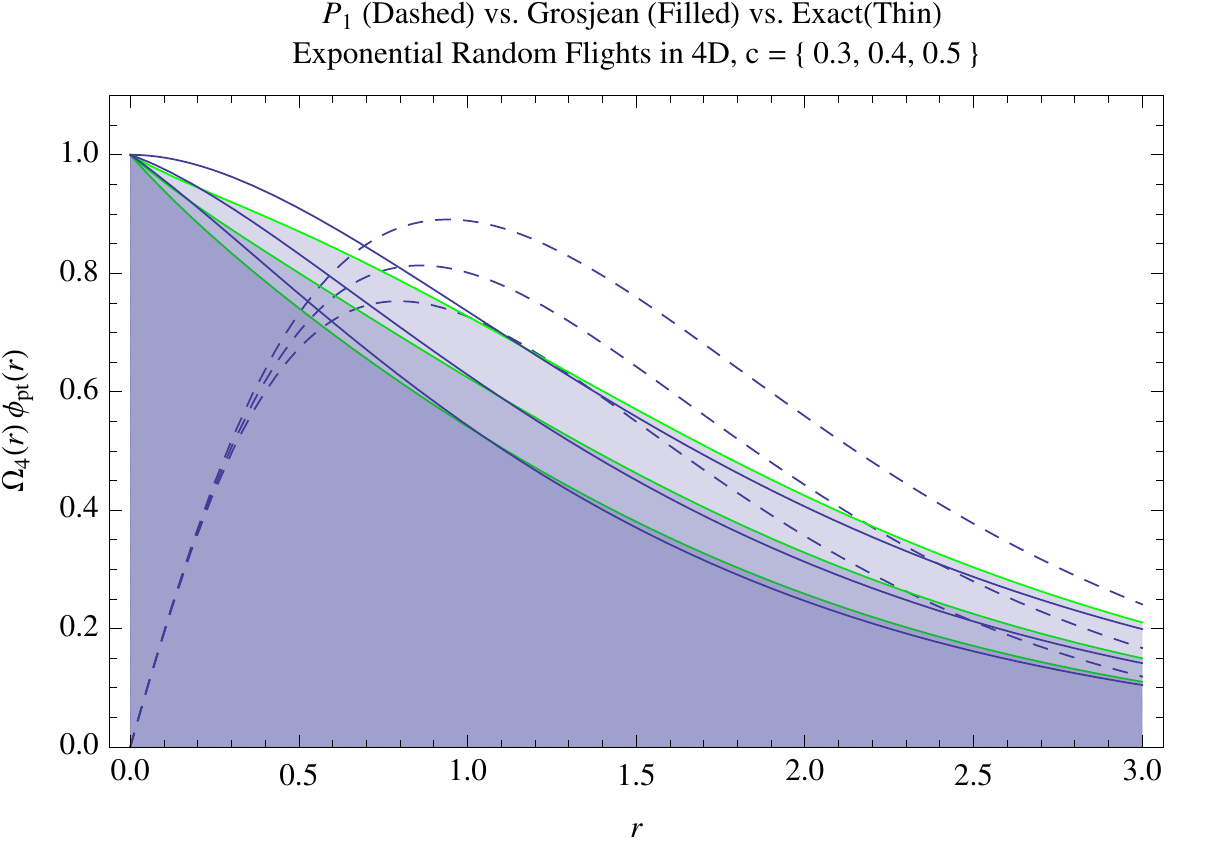}}
        \subfigure[High absorption - relative error]{\includegraphics[width=.48\linewidth]{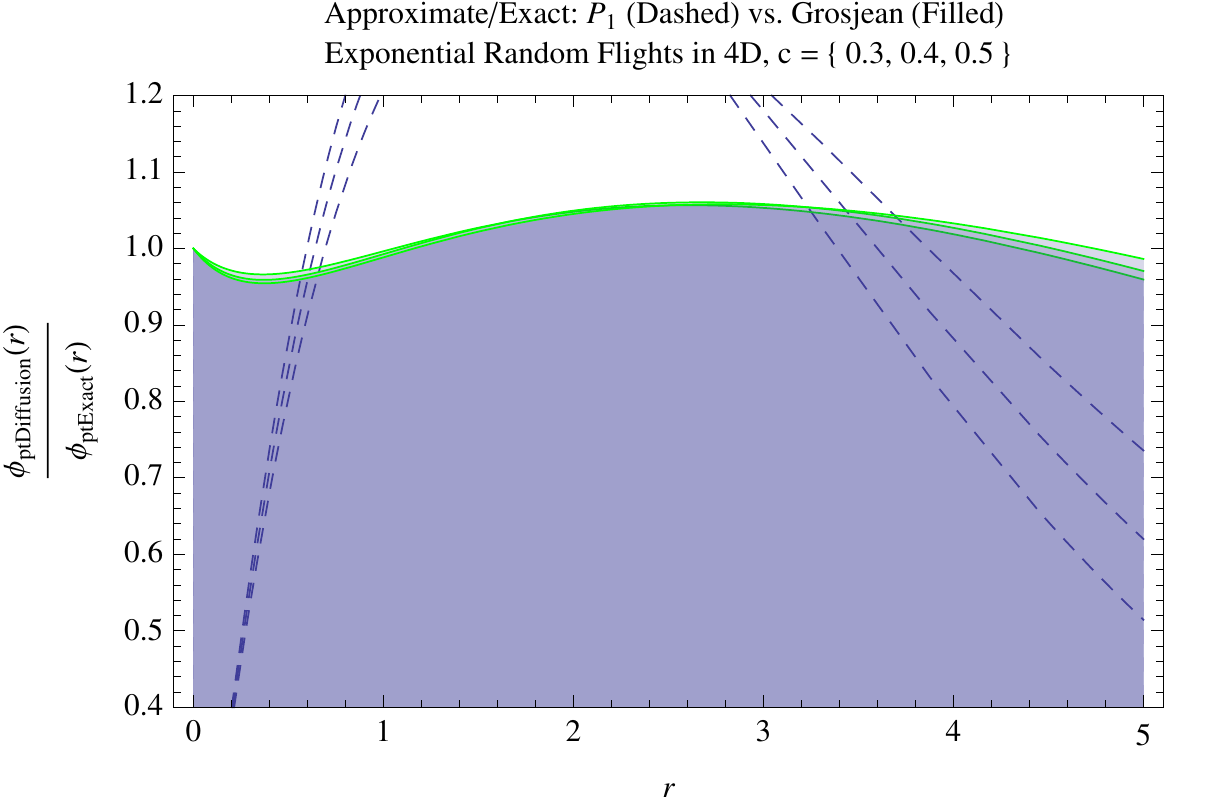}}
        \caption{Comparison of the accuracy of classical ($P_1$), rigorous asymptotic and Grosjean modified diffusion approximations for the scalar flux $\phipt(r)$ about an isotropic point source in an infinite 4D medium.}
        \label{fig-expdiffusionCompare4D} 
      \end{figure*}

      The \changed{Pad\'{e}} approximant of the total transformed scalar flux
      \begin{equation}
        \frac{2}{1-2 c+\sqrt{z^2+1}} \approx \frac{4}{-4 c+z^2+4}
      \end{equation}
      produces the classical $P_1$ diffusion approximation in 4D
      \begin{equation}
        \phipt(r) \approx \frac{2 \sqrt{1-c} K_1\left(2 \sqrt{1-c} r\right)}{\pi ^2 r}.
      \end{equation}
      The \changed{Pad\'{e}} approximant of the transformed scattered scalar flux
      \begin{equation}
        \frac{4 c \left(\sqrt{z^2+1}-1\right)^2}{z^2 \left(z^2-2 c \left(\sqrt{z^2+1}-1\right)\right)} \approx \frac{4 c}{(2-c) z^2+4 (1-c)}
      \end{equation}
      produces the Grosjean modified-diffusion approximation in 4D
      \begin{equation}
        \phipt(r) \approx \frac{\e^{-r}}{2 \pi^2 r^3} + \frac{2 \sqrt{\frac{1}{c-2}+1} c K_1\left(2 \sqrt{1+\frac{1}{c-2}}
   r\right)}{\pi ^2 (2-c) r}.
      \end{equation}
      Figure~\ref{fig-expdiffusionCompare4D} compares the various diffusion approximations for exponential random flights in 4D.  Similar to 3D, we see the Grosjean and rigorous diffusion approximations outperforming $P_1$.  In contrast to flatland, where the performance of Grosjean vs. exact lessens for high $c$, in 4D, Grosjean diffusion is remarkably accurate.  The rigorous diffusion approximations are not included in the high absorption plots, because the approximation breaks down.
      
  \subsection{Exponential---General Dimension}
    \subsubsection{Propagators}
      For multiple scattering with an exponential free path distribution in a general, $d$-dimensional medium, the transformed propagator is~\cite{zoia11e}
      \begin{equation}
        \o{\zeta}_d(z) = \, _2F_1\left(\frac{1}{2},1;\frac{d}{2};-z^2\right),
      \end{equation}
      where $\, _2F_1$ is a hypergeometric function.   Propagators for the first few integer dimensions are
      \begin{equation*}
      \begin{array}{ll}
       d & \o{\zeta}_d(z) \\
       1 & \frac{1}{z^2+1} \\
       2 & \frac{1}{\sqrt{z^2+1}} \\
       3 & \frac{\arctan z}{z} \\
       4 & \frac{2 \left(\sqrt{z^2+1}-1\right)}{z^2} \\
       5 & \frac{3 \left(z^2 \arctan z-z+\arctan z\right)}{2 z^3} \\
       6 & \frac{4 \left(2 \sqrt{z^2+1} z^2-3 z^2+2 \sqrt{z^2+1}-2\right)}{3 z^4}. \\
      \end{array}
      \end{equation*}

    \subsubsection{Collision Densities (Exponential, General $d$)}
      For exponential random flights in $d$ dimensions, the first few even collision density moments are
      \changedcomment{Fixed $\Omega_d$ instead of $\Omega_4$}{
      \begin{align}
        &\int_0^\infty \Omega_d(r) \Cpt(r|n) dr = c^{n-1} \\
        &\int_0^\infty \Omega_d(r) r^2 \Cpt(r|n) dr = 2 c^{n-1} n \\
        &\int_0^\infty \Omega_d(r) r^4 \Cpt(r|n) dr = \frac{4 n c^{n-1} ((d+2) n+5 d-2)}{d} \\
        &\int_0^\infty \Omega_d(r) r^6 \Cpt(r|n) dr = \\ &\frac{8 n c^{n-1} \left(d^2 (n (n+15)+74)+6 d (n-1) (n+10)+8 (n-2) (n-1)\right)}{d^2}
      \end{align}
      and the total collision density moments
      \begin{align}
        &\int_0^\infty \Omega_d(r) \Cpt(r) dr = \frac{1}{1-c} \\
        &\int_0^\infty \Omega_d(r) r^2 \Cpt(r) dr = \frac{2}{(c-1)^2} \\
        &\int_0^\infty \Omega_d(r) r^4 \Cpt(r) dr = \frac{8 (2 c (d-1)-3 d)}{(c-1)^3 d} \\
        &\int_0^\infty \Omega_d(r) r^6 \Cpt(r) dr = \\ &\frac{48 \left(2 c^2 (d-1) (5 d-4)-24 c (d-1) d+15 d^2\right)}{(c-1)^4 d^2}.
      \end{align}
      }
      Here we note that the zeroth and second spatial moments of the distribution are independent of dimension $d$, whereas the higher order moments do depend on $d$.

    \subsubsection{Diffusion Approximations (Exponential, general $d$)}
      The \changed{Pad\'{e}} approximant for the scalar flux / fluence
      \begin{equation}
        \frac{\o{X}(z)}{1 - \ssalbedo \o{\zeta}_d(z)} \approx \frac{d}{ d( 1 - c ) + z^2 }
      \end{equation}
      produces the classical $P_1$ diffusion length (in mean free paths) of $1 / \sqrt{d(1-c)}$.  This proves the conjecture by Asadzadeh and Larsen~\shortcite{asadzadeh08} that the appearance of ``$3$" in the diffusion coefficient in 3D, changing to ``$2$" in flatland, and ``$1$" in the rod indeed extends to any positive dimensional space with classical transport.

      The rigorous diffusion approximation is the sum of all diffusion terms
      \begin{multline}
        \phipt(r) \approx \\ \sum_\chi \frac{d}{c^2 \chi^2 \, _2F_1\left(\frac{3}{2},2;\frac{d}{2}+1;\chi^2\right)} (2 \pi )^{-d/2} r^{1-\frac{d}{2}} (\frac{1}{\chi})^{-\frac{d}{2}-1}
     K_{\frac{d-2}{2}}\left(r \chi \right) 
      \end{multline}
      \changed{where the sum is over all real}, positive root $\chi > 1$ of the characteristic equation
      \begin{equation}
          1 - c \, _2F_1\left(\frac{1}{2},1;\frac{d}{2};\chi^2\right) = 0.
      \end{equation}
      In three or fewer dimensions with isotropic scattering there appears to always be a single discrete eigenvalue $\nu_0 > 1$.  However, this trend does not continue to higher integer dimensions.  In 4 dimensions, we have seen that the discrete eigenvalue goes to $1$ for $c = 0.5$ and the rigorous diffusion mode disappears (and simultaneously, for this unique absorption level, we are able to invert the transformed collision density and produce an exact solution).  Interestingly, in 6D the solutions of the characteristic equation are
      \begin{equation}
        \nu_0 = 
        \begin{cases}
          \pm  \left( \frac{2}{3} \sqrt{(9-8 c) c-\sqrt{c (4 c-3)^3}} \right)^{-1} \\
          \pm  \left( \frac{2}{3} \sqrt{(9-8 c) c+\sqrt{c (4 c-3)^3}} \right)^{-1}
        \end{cases} 
      \end{equation}
      and we see two positive, real discrete rigorous diffusion eigenmodes when $c > 0.75$.

      The \changed{Pad\'{e}} approximant for the collided scalar flux / fluence
      \begin{equation}
        \frac{\o{X}(z) \ssalbedo \o{\zeta}_d(z)}{1 - \ssalbedo \o{\zeta}_d(z)} \approx \frac{c d}{(1-c) d+(2-c) z^2}
      \end{equation}
      produces the modified Grosjean diffusion length (in mean free paths) of
      \begin{equation}
        \nu_0 = \sqrt{\frac{2-c}{(1-c) d}}.
      \end{equation}
      Exactly analagous to the $P_1$ diffusion length we see the same $1 / \sqrt{d}$ dependence on the Grosjean diffusion length.  Considering, for the moment, a general mean-free path $\ell > 0$, the Grosjean diffusion approximation for the scalar flux about a point source in $d$ dimensions is
      \begin{equation}
        \phipt(r) \approx \frac{E(r)}{\Omega_d(r)} + \frac{c \, r \left(\frac{c-2 \ell^2}{c d \ell^2-d
     \ell^4}\right)^{\frac{1}{4} (-d-2)} K_{\frac{d-2}{2}}\left(\frac{d \ell r}{\sqrt{d
     \left(\frac{c}{\ell^2-c}+2\right)}}\right)}{(2 \pi r )^{d/2} (\ell^3-c \ell)}.
      \end{equation}

\begin{figure*}
        \centering
        \subfigure[$c = 0.75$]{\includegraphics[width=.33\linewidth]{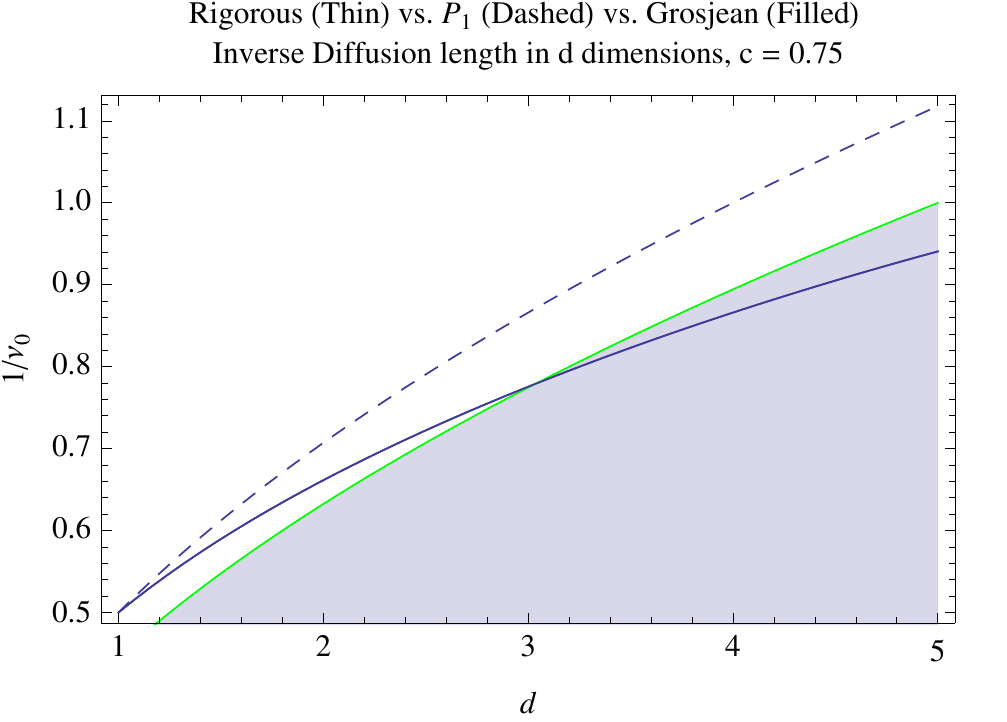}}
        \subfigure[$c = 0.9$]{\includegraphics[width=.33\linewidth]{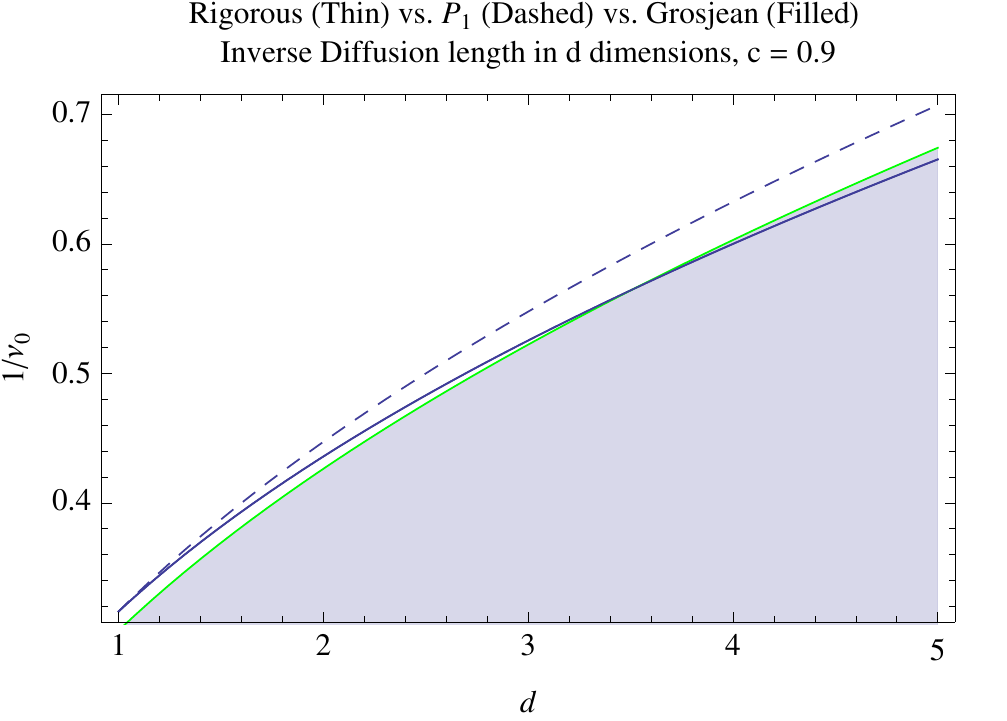}}
        \subfigure[$c = 0.99$]{\includegraphics[width=.33\linewidth]{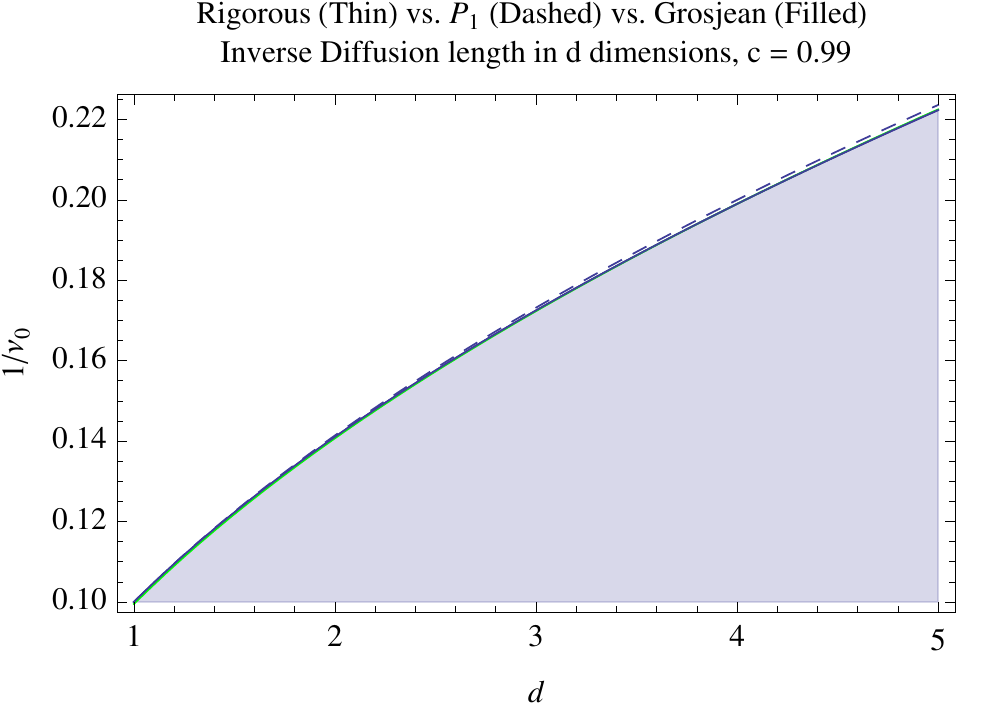}}
        \caption{Comparison of the diffusion \changed{lengths} predicted by classical $P_1$, rigorous asymptotic, and Grosjean modified diffusion theory for various dimensions and absorption levels.}
        \label{fig-expv0compare} 
      \end{figure*}

      Figure~\ref{fig-expv0compare} compares the diffusion lengths predicted by the various diffusion theories as a function of dimension for select levels of absorption.  The rigorous diffusion lengths are solved for using numerical root finding in MATHEMATICA.  We see a clear trend of decreasing diffusion length with dimension in all cases (the inverse diffusion length increases in the plots).  \changedcomment{wording change}{The $P_1$ and Grosjean diffusion lengths both} have a $1 / \sqrt{d}$ dependence on dimension.  We see this is a fairly close match for rigorous diffusion lengths for very low absorption, but as absorption increases we see the rigorous diffusion lengths decreasing faster than $1 / \sqrt{d}$.

\section{Generalized Linear Transport theory (non-exponential random flights)}\label{sec:flights}

In the following sections we derive new diffusion approximations for a variety of generalized linear transport theories by considering several classes of common continuous distributions for the free path distribution function $p(s)$.  When a chosen free path distribution has an unbounded mean or square mean (for example, a Cauchy or Levy distribution), \emph{anomalous-}, as opposed to classical-, diffusion arises~\cite{metzler04}, which we do not consider presently.  It is also possible that the free path distribution has a finite square mean but not a finite mean square extinction (such as the Beta distribution $\frac{16 \sqrt{s}}{\pi  (s+1)^4}$).  In this case, diffusion theory can only describe the collision density, and not the scalar flux / fluence.  Indeed, we find in the following generalized transport theories that the classical and Grosjean diffusion approximations are distinctly different for the collision densities vs. the scalar flux / fluence distributions.

\begin{figure*}
	\centering
	\subfigure[Beta-prime]{\includegraphics[width=.3\linewidth]{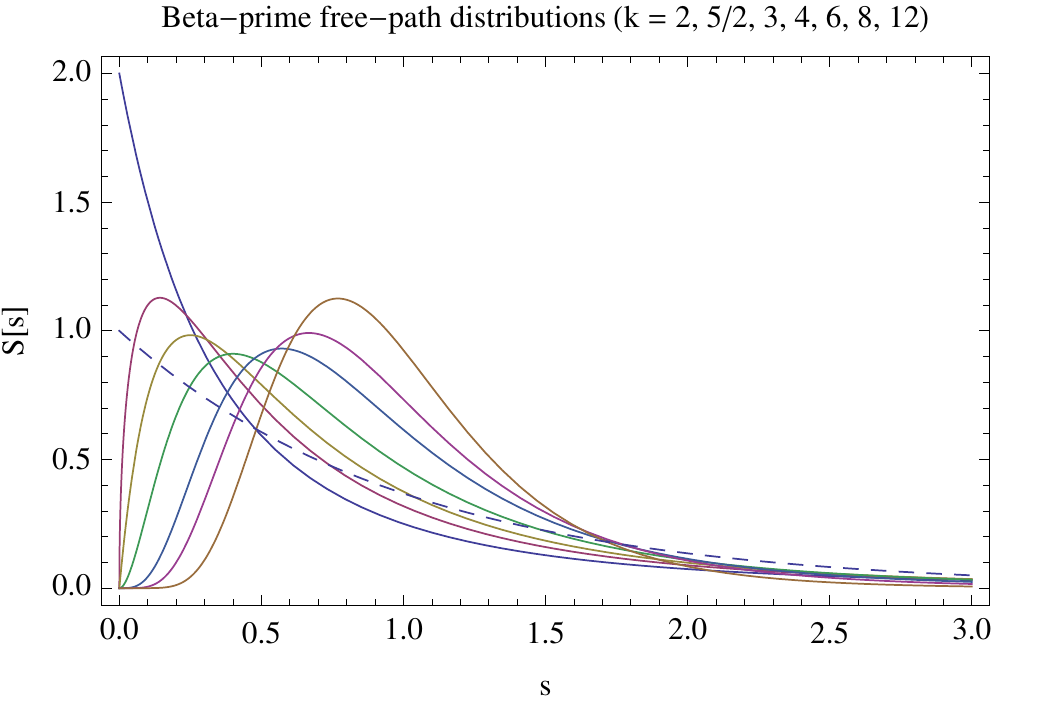}}
	\subfigure[Chi]{\includegraphics[width=.3\linewidth]{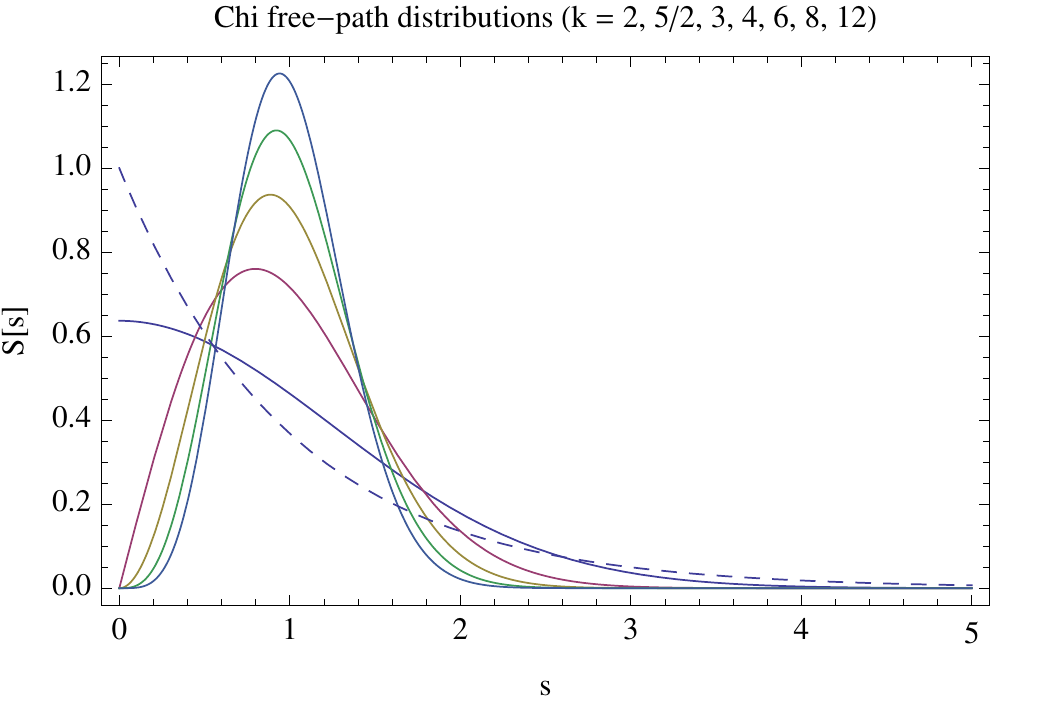}}
	\subfigure[Gamma]{\includegraphics[width=.3\linewidth]{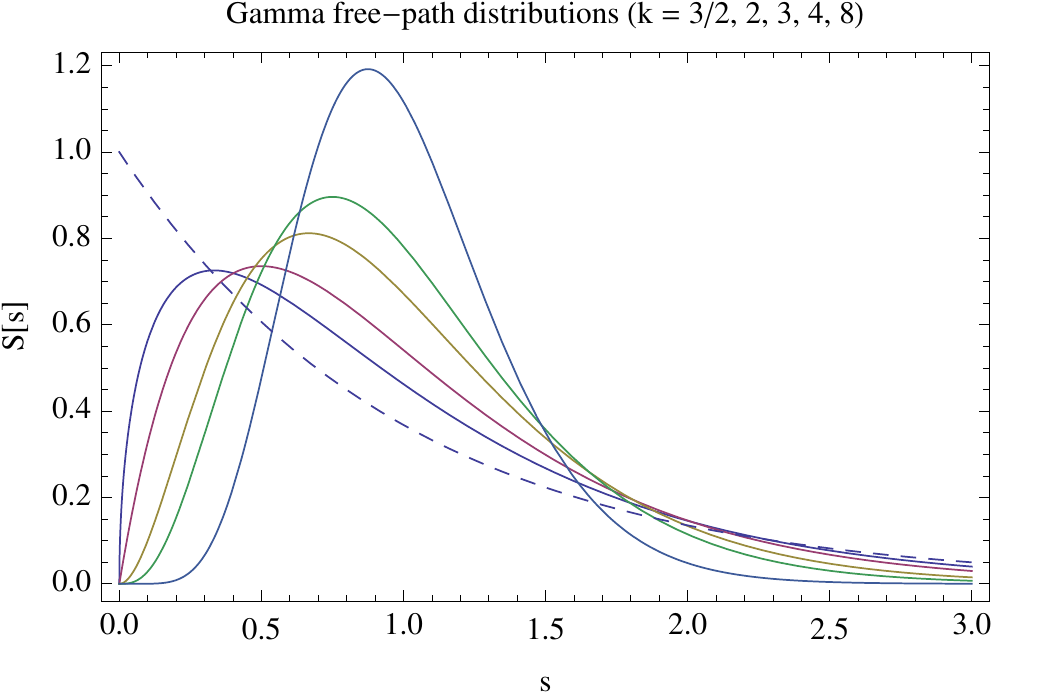}}
	\caption{Examples of the variety of mean free path distributions for which we derive diffusion approximations in this paper.  All distributions have a unit mean-free path $\ell = 1$.  All figures compare to the exponential free path distribution of classical transport theory/radiative transfer (dashed).}
	\label{fig-meanfreepaths} 
\end{figure*}

Figure~\ref{fig-meanfreepaths} compares the various families of free path distributions that we consider.  All plots show distributions with a unit mean free path ($\ell = 1$) and with the classical exponential distribution shown in each plot for reference.

\section{Beta-prime random flights}
    Here we consider non-classical transport with a beta-prime free path distribution with unit mean and parameter $k > 2$,
    \begin{equation}
      p(s) = \frac{s^{k-2} (s+1)^{1-2 k}}{B(k-1,k)},
    \end{equation}
    where $B$ is the beta function.  Examples include
    \begin{equation}
      \left\{\frac{16 \sqrt{s}}{\pi  (s+1)^4},\frac{12 s}{(s+1)^5},\frac{60 s^2}{(s+1)^7},\frac{24024 s^6}{(s+1)^{15}}\right\} \in p(s).
    \end{equation}
    The beta-prime extinction is
    \begin{equation}
      E(s) = \frac{(-1)^{-k} B_{-s}(k-1,2-2 k)}{B(k-1,k)}+1,
    \end{equation}
    the mean square free path is
    \begin{equation}
      \langle s^2 \rangle = \frac{k}{k-2},
    \end{equation}
    and the mean square extinction is
    \begin{equation}
      \int_0^\infty E(s) s^2 = \frac{k (k+1)}{3 \left(k^2-5 k+6\right)} \, \, \, ( k > 3 ).
    \end{equation}

  \subsection{General Dimension (Beta-prime)}
    \subsubsection{Diffusion Approximations (Beta-prime, General $d$)}

    The transformed propagators for beta-prime flights can be found in MATHEMATICA, but are very complex expressions of MeijerG functions and FresnelS functions.  We were unable to inverse transform any multiple collision densities, nor find any rigorous diffusion lengths in closed form.

    We conjecture that the classical $P_1$ diffusion approximation for the beta-prime collision density in $d$ dimensions is
    \begin{equation}
      \Cpt(r) \approx \frac{1}{1-c} \mathcal{F}_3^{-1} \left\{ \frac{1}{1+(z v_0)^2} \right\}
    \end{equation}
    with diffusion length
    \begin{equation}
      \nu_0 = \frac{1}{\sqrt{2} \sqrt{\frac{(1-c) d (k-2)}{k}}}.
    \end{equation}

    Similarly, we conjecture that the Grosjean diffusion length for approximating the collision density is
    \begin{equation}
      \frac{1}{v_G} = \sqrt{2} \sqrt{\frac{(c-1) d (k-2)}{(c-2) k}},
    \end{equation}
    which we were able to confirm in 3D for values of $k \in \{ \frac{5}{2}, \frac{7}{2}, \frac{9}{2}, ..., \frac{23}{2} \}$.
    
 \section{Chi Random Flights}

  Here we consider non-classical transport with a Chi free path distribution with unit mean and parameter $k \ge 1$,
    \begin{equation}
      p(s) = 2 s^{k-1} \Gamma \left(\frac{k}{2}\right)^{-k-1} \Gamma \left(\frac{k+1}{2}\right)^k e^{-\frac{s^2 \Gamma \left(\frac{k+1}{2}\right)^2}{\Gamma \left(\frac{k}{2}\right)^2}}.
    \end{equation}
    Examples include
    \begin{equation}
      \left\{\frac{2 e^{-\frac{s^2}{\pi }}}{\pi },\frac{1}{2} \pi  e^{-\frac{\pi  s^2}{4}} s,\frac{32 e^{-\frac{4 s^2}{\pi }}
   s^2}{\pi ^2},\frac{81}{128} \pi ^2 e^{-\frac{9 \pi  s^2}{16}} s^3\right\} \in p(s),
    \end{equation}
    the first example with $k = 1$ being the half-normal distribution.  The extinction function for Chi flights is
    \begin{equation}
      E(s) = \frac{\Gamma \left(\frac{k}{2},\frac{s^2 \Gamma \left(\frac{k+1}{2}\right)^2}{\Gamma
   \left(\frac{k}{2}\right)^2}\right)}{\Gamma \left(\frac{k}{2}\right)},
    \end{equation}
    the mean square free path is
    \begin{equation}
      \langle s^2 \rangle = \frac{\Gamma \left(\frac{k}{2}+1\right) \Gamma \left(\frac{k}{2}\right)}{\Gamma
   \left(\frac{k+1}{2}\right)^2}
    \end{equation}
    and the mean square extinction is
    \begin{equation}
      \int_0^\infty E(s) s^2 = \frac{(k+1) \Gamma \left(\frac{k}{2}\right)^2}{6 \Gamma \left(\frac{k+1}{2}\right)^2}.
    \end{equation}
    When $k = d$, the Chi free-path distribution produces Gaussian $n$th collision densities.

  \subsection{Chi Random Flights---Rod Model ($d = 1$)}
    \subsubsection{Rod Model - Chi $k = 1$}
      For $k = 1$, the transformed propagator in the rod is
      \begin{equation}
        \o{\zeta_1}(z) = e^{-\frac{\pi  z^2}{4}}.
      \end{equation}
      The characteristic equation yields single positive discrete eigenvalues for $c > 0$,
      \begin{equation}
        \nu_0 = \frac{\sqrt{\pi }}{2 \sqrt{\log \left(\frac{1}{c}\right)}}.
      \end{equation}
      The $n$th collision density is the repeated convolution of a Gaussian propagator and is also, trivially, Gaussian
      \begin{equation}
        \Cpt(r|n) = \frac{ c^{n-1} e^{-\frac{r^2}{\pi  n}}}{\pi  \sqrt{n}}.
      \end{equation}
      The $n$th collision moments of order $m$ are
      \begin{equation}
        \int_0^\infty \Omega_1(r) r^m \Cpt(r|n) dr = \pi ^{\frac{m-1}{2}} c^{n-1} n^{m/2} \Gamma \left(\frac{m+1}{2}\right)
      \end{equation}
      and sum to give the total collision density moments
      \begin{equation}
        \int_0^\infty \Omega_1(r) r^m \Cpt(r) dr = \frac{\pi ^{\frac{m-1}{2}} \Gamma \left(\frac{m+1}{2}\right) \text{Li}_{-\frac{m}{2}}(c)}{c},
      \end{equation}
      where $\text{Li}_n(x)$ is the polylogarithm function.

    \subsubsection{Rod Model - Chi $k = 3$}
      For $k = 3$, the transformed propagator in the rod is
      \begin{equation}
        \o{\zeta_1}(z) = -\frac{1}{8} e^{-\frac{\pi  z^2}{16}} \left(\pi  z^2-8\right).
      \end{equation}
      The second collision density can be solved for explicitly,
      \begin{equation}
        \Cpt(r|2) = \frac{c e^{-\frac{2 r^2}{\pi }} \left(16 r^4-8 \pi  r^2+3 \pi
   ^2\right)}{2 \sqrt{2} \pi ^3}.
      \end{equation}

  \subsection{Chi Random Flights---Flatland ($d = 2$)}
    \subsubsection{Flatland - Chi $k = 2$}
      For $k = 2$, the transformed propagator in flatland is
      \begin{equation}
        \o{\zeta_2}(z) = e^{-\frac{z^2}{\pi }}.
      \end{equation}
      The characteristic equation yields single positive discrete eigenvalues for $c > 0$,
      \begin{equation}
        \nu_0 = \frac{1}{\sqrt{\pi } \sqrt{\log \left(\frac{1}{c}\right)}}
      \end{equation}
      The $n$th collision density is the repeated convolution of a Gaussian propagator and is also, trivially, Gaussian
      \begin{equation}
        \Cpt(r|n) = \frac{c^{n-1} e^{-\frac{\pi  r^2}{4 n}}}{4 n}
      \end{equation}

\begin{figure*}
        \centering
        \subfigure[Low absorption]{\includegraphics[width=.9\linewidth]{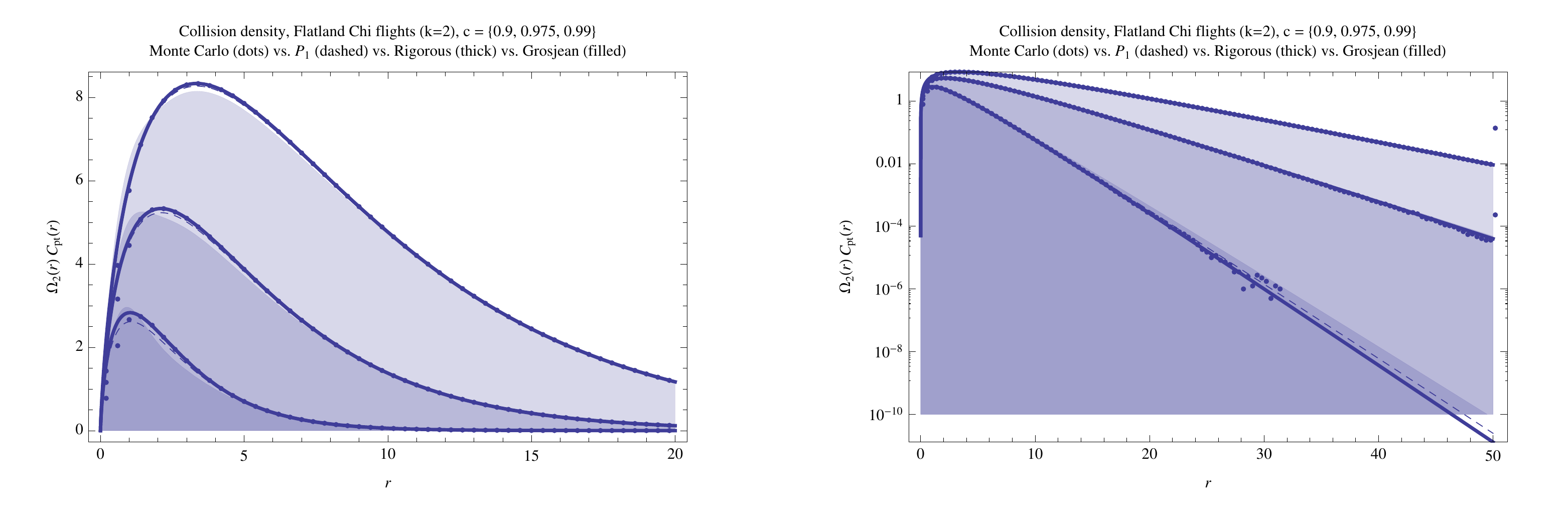}}
        \subfigure[High absorption]{\includegraphics[width=.9\linewidth]{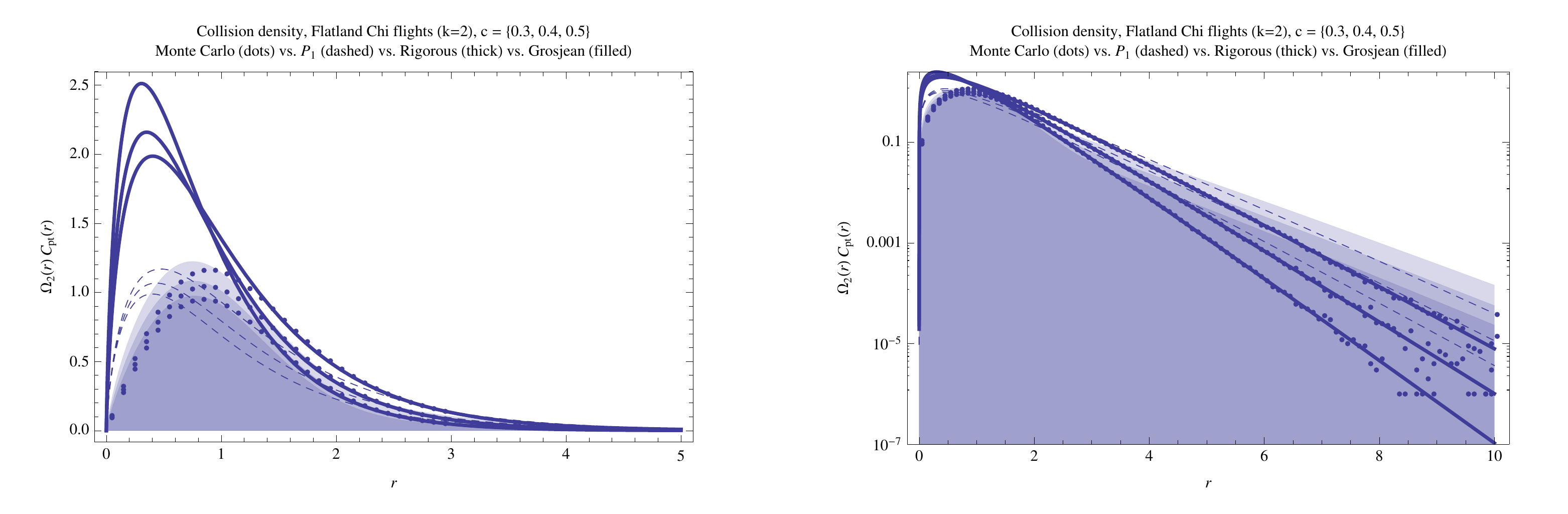}}
        \caption{Non classical linear Boltzmann transport in flatland with Chi-distributed free paths.  Similar to classical transport in three dimensions, we see Grosjean and rigorous asymptotic diffusion approximations of the collision density outperforming the classical $P_1$ diffusion approximation.}
        \label{fig-chik2flatlanddiffusion} 
      \end{figure*}

      The \changed{Pad\'{e}} approximant of the total transformed collision density
      \begin{equation}
        \frac{1}{e^{\frac{z^2}{\pi }}-c} \approx \frac{\pi }{-\pi  c+z^2+\pi }
      \end{equation}
      whose inversion produces the classical $P_1$ diffusion approximation
      \begin{equation}
        \Cpt(r) \approx \frac{1}{2} K_0\left(\sqrt{\pi -c \pi } r\right)
      \end{equation}
      The Pade\'{e} approximant of the transformed multiply-scattered collision density
      \begin{equation}
        \frac{c e^{-\frac{z^2}{\pi }}}{e^{\frac{z^2}{\pi }}-c} \approx -\frac{\pi  c}{(c-2) z^2+\pi  (c-1)}
      \end{equation}
      leads to the Grosjean modified-diffusion approximation for the collision density
      \begin{equation}
        \Cpt(r) \approx \frac{e^{-\frac{\pi  s^2}{4}} s}{4 r} +\frac{c K_0\left(\sqrt{1+\frac{1}{c-2}} \sqrt{\pi } r\right)}{2 (2-c)}.
      \end{equation}
      The rigorous asymptotic diffusion approximation for the collision density for Chi flights in flatland with $k = 2$ is
      \begin{equation}
        \Cpt(r) \approx \frac{K_0\left(\sqrt{\pi } r \sqrt{\log \left(\frac{1}{c}\right)}\right)}{2 c}.
      \end{equation}
      Figure~\ref{fig-chik2flatlanddiffusion} compares the three diffusion approximations to Monte Carlo reference solutions for the collision density about the point source.  Similar to classical transport in three dimensions we see the rigorous asymptotic diffusion approximation achieving the highest accuracy for low densities, far from the source where the transient terms are negligible.  However, unlike classical transport theory, here we see the rigorous asymptotic diffusion approximation overpredicting the result near the source, instead of underpredicting it.  Also, just as in classical transport, we see Grosjean's modified diffusion approximation performing the best near the source for high absorption levels.  Very low absorption leads to broader, improved accuracy for all three approximations, and they become quite similar to each other.

    \subsubsection{Flatland - Chi $k = 4$}
      For $k = 4$, the transformed propagator in flatland is
      \begin{equation}
        \o{\zeta_2}(z) =\frac{e^{-\frac{4 z^2}{9 \pi }} \left(9 \pi -4 z^2\right)}{9 \pi }.
      \end{equation}
      The second collision density can be solved for explicitly,
      \begin{equation}
        \Cpt(r|2) = \frac{9 c e^{-\frac{9 \pi  r^2}{32}} \left(81 \pi ^2
   r^4+2048\right)}{131072}
      \end{equation}

  \subsection{Chi Random Flights---3D ($d = 3$)}
    \subsubsection{Chi - 3D ($k = 3$)}
      For $k = 3$, the transformed propagator in 3D is
      \begin{equation}
        \o{\zeta_3}(z) = e^{-\frac{\pi  z^2}{16}}
      \end{equation}
      The characteristic equation yields single positive discrete eigenvalues for $c > 0$,
      \begin{equation}
        \nu_0 = \frac{\sqrt{\pi }}{4 \sqrt{\log \left(\frac{1}{c}\right)}}
      \end{equation}
      The $n$th collision density is the repeated convolution of a Gaussian propagator and is also, trivially, Gaussian
      \begin{equation}
        \Cpt(r|n) = \frac{8 c^{n-1} e^{-\frac{4 r^2}{\pi n}} \sqrt{\frac{1}{n^3}}}{\pi ^3
   }.
      \end{equation}

      The classical $P_1$ diffusion approximation for the collision density is
      \begin{equation}
        \Cpt(r) \approx \frac{4 e^{-\frac{4 \sqrt{1-c} r}{\sqrt{\pi }}}}{\pi ^2 r}.
      \end{equation}
      The Grosjean modified-diffusion approximation for the collision density is
      \begin{equation}
        \Cpt(r) \approx \frac{8 e^{-\frac{4 r^2}{\pi }} r^2}{\pi ^3 r^2}+\frac{4 c e^{-\frac{4 \sqrt{\frac{c-1}{c-2}}
   r}{\sqrt{\pi }}}}{\pi ^2 (2-c) r}.
      \end{equation}
      The rigorous asymptotic diffusion approximation for the collision density is
      \begin{equation}
        \Cpt(r) \approx \frac{4 e^{-\frac{4 r \sqrt{-\log (c)}}{\sqrt{\pi }}}}{\pi ^2 c r}.
      \end{equation}

  \subsection{Chi Random Flights---General Dimension}
    For multiple scattering in a general, $d$-dimensional medium, the transformed Chi propagator is
    \begin{equation}
      \o{\zeta_d}(z) = \, _1F_1\left(\frac{k}{2};\frac{d}{2};-\frac{z^2 \Gamma \left(\frac{k}{2}\right)^2}{4 \Gamma
     \left(\frac{k+1}{2}\right)^2}\right)
    \end{equation}
    where $\, _1F_1$ is a hypergeometric function.

    Chi random flights appear to always have a single real discrete eigenvalue, which is approximated well for $c > 0.8$ by approximating the transformed propagator with a $(2,2)$ \changed{Pad\'{e}} approximant and solving the characteristic equation, yielding
    \begin{equation}
      \frac{1}{\nu_0} \approx \frac{2 \sqrt{2} \sqrt{c-1} \sqrt{d^2+2 d} \Gamma \left(\frac{k+1}{2}\right)}{\sqrt{2 c d \Gamma
   \left(\frac{k}{2}\right)^2-c d k \Gamma \left(\frac{k}{2}\right)^2-4 c k \Gamma \left(\frac{k}{2}\right)^2-2 d \Gamma
   \left(\frac{k}{2}\right)^2-d k \Gamma \left(\frac{k}{2}\right)^2}}.
    \end{equation}

    For Chi random flights in $d$ dimensions, the first few even collision density moments are
      \begin{align}
        &\int_0^\infty \Omega_d(r) \Cpt(r|n) dr = c^{n-1} \\
        &\int_0^\infty \Omega_d(r) r^2 \Cpt(r|n) dr = \frac{k n c^{n-1} \Gamma \left(\frac{k}{2}\right)^2}{2 \Gamma \left(\frac{k+1}{2}\right)^2} \\
        &\int_0^\infty \Omega_d(r) r^4 \Cpt(r|n) dr = \frac{k n c^{n-1} \Gamma \left(\frac{k}{2}\right)^4 (d (k n+2)+2 k (n-1))}{4 d \Gamma
   \left(\frac{k+1}{2}\right)^4}
      \end{align}
      and the total collision density moments
      \begin{align}
        &\int_0^\infty \Omega_d(r) \Cpt(r) dr = \frac{1}{1-c} \\
        &\int_0^\infty \Omega_d(r) r^2 \Cpt(r) dr = \frac{k \Gamma \left(\frac{k}{2}\right)^2}{2 (c-1)^2 \Gamma \left(\frac{k+1}{2}\right)^2} \\
        &\int_0^\infty \Omega_d(r) r^4 \Cpt(r) dr = -\frac{k \Gamma \left(\frac{k}{2}\right)^4 (d (c (k-2)+k+2)+4 c k)}{4 (c-1)^3 d \Gamma
   \left(\frac{k+1}{2}\right)^4}.
      \end{align}

      The first few even $n$th scattered scalar flux / fluence moments are
      \begin{align}
        &\int_0^\infty \Omega_d(r) \phipt(r|n) dr = c^n \\
        &\int_0^\infty \Omega_d(r) r^2 \phipt(r|n) dr = \frac{c^n (3 k n+k+1) \Gamma \left(\frac{k}{2}\right)^2}{6 \Gamma
   \left(\frac{k+1}{2}\right)^2} \\
        &\int_0^\infty \Omega_d(r) r^4 \phipt(r|n) dr = \\ &\frac{c^n \Gamma \left(\frac{k}{2}\right)^4 \left(15 (d+2) k^2 n^2+10 k n (d (k+4)-k+2)+3 d
   (k+1) (k+3)\right)}{60 d \Gamma \left(\frac{k+1}{2}\right)^4}
      \end{align}
      and the total scalar flux / fluence moments are
      \begin{align}
        &\int_0^\infty \Omega_d(r) \phipt(r) dr = \frac{1}{1-c} \\
        &\int_0^\infty \Omega_d(r) r^2 \phipt(r) dr = \frac{(c (2 k-1)+k+1) \Gamma \left(\frac{k}{2}\right)^2}{6 (c-1)^2 \Gamma
   \left(\frac{k+1}{2}\right)^2}.
      \end{align}
      Again, we see that the first two even moments are dimension independent.

    \subsubsection{Chi Flights Collision Density Diffusion}
      The Pade\'{e} approximant producing the transformed classical $P_1$ diffusion approximation for the total collision density is
      \begin{equation}
        \frac{\o{\zeta}_d(z)}{1 - \ssalbedo \o{\zeta}_d(z)} \approx \frac{1}{-c+\frac{k z^2 \Gamma \left(\frac{k}{2}\right)^2}{4 d \Gamma
     \left(\frac{k+1}{2}\right)^2}+1}
      \end{equation}
      which supports our conjecture (Equation~\ref{eq:P1Conjecture}) for all dimensions $d$ and parameter $k$.  Similarly, the approximant for Grosjean's diffusion approximation for Chi flights is
      \begin{equation}
        \frac{c (\o{\zeta}_d(z))^2}{1 - \ssalbedo \o{\zeta}_d(z)} \approx -\frac{4 c d \Gamma \left(\frac{k+1}{2}\right)^2}{4 (c-1) d \Gamma
   \left(\frac{k+1}{2}\right)^2+(c-2) k z^2 \Gamma \left(\frac{k}{2}\right)^2}
      \end{equation}
      and likewise supports our second conjecture (Equation~\ref{eq:grosjeanConjecture}).

    \subsubsection{Chi Flights Scalar Flux Diffusion}
      We derive the classical $P_1$ diffusion approximation for the scalar flux / fluence for Chi random flights from the Pade\'{e} approximant
      \begin{equation}
        \frac{\o{X}(z)}{1 - \ssalbedo \zeta_d(z)} \approx \frac{1}{\frac{z^2 (c (2 k-1)+k+1) \Gamma \left(\frac{k}{2}\right)^2}{12 d \Gamma \left(\frac{k+1}{2}\right)^2}-c+1}
      \end{equation}
      yielding the scalar flux diffusion length
      \begin{equation}
        \nu_0^2 = \frac{(c (2 k-1)+k+1) \Gamma \left(\frac{k}{2}\right)^2}{12 (1-c) d \Gamma
   \left(\frac{k+1}{2}\right)^2}.
      \end{equation} 
      Similarly, Grosjean's modified diffusion approximation for the scalar flux / fluence for Chi random flights derives from the \changed{Pad\'{e}} approximant
      \begin{multline}
        \frac{\o{X}(z) c \o{\zeta}_d(z)}{1 - \ssalbedo \zeta_d(z)} \approx \\ -\frac{12 c d \Gamma \left(\frac{k+1}{2}\right)^2}{12 (c-1) d \Gamma
   \left(\frac{k+1}{2}\right)^2+z^2 ((c-4) k+c-1) \Gamma \left(\frac{k}{2}\right)^2}
      \end{multline}
      yielding the scalar flux Grosjean diffusion length
      \begin{equation}
        v_G^2 = \frac{((c-4) k+c-1) \Gamma \left(\frac{k}{2}\right)^2}{12 (c-1) d \Gamma
   \left(\frac{k+1}{2}\right)^2}.
      \end{equation}

 \section{Gamma (Erlang) Random Flights}

  Here we consider non-classical transport with a Gamma free path distribution with unit mean and parameter $k > 0$,
    \begin{equation}
      p(s) = \frac{\left(\frac{1}{k}\right)^{-k} e^{-k s} s^{k-1}}{\Gamma (k)}.
    \end{equation}
    When $k$ is an integer, this corresponds to the Erlang distribution, and for $k \rightarrow k / 2$, becomes the Chi-squared distribution.  Examples include
    \begin{equation}
      \left\{\frac{e^{-s/2}}{\sqrt{2 \pi } \sqrt{s}},e^{-s},4 e^{-2 s} s,\frac{27}{2}
   e^{-3 s} s^2,\frac{128}{3} e^{-4 s} s^3\right\} \in p(s),
    \end{equation}
    the second example with $k = 1$ being the exponential distribution.  The extinction function for Gamma flights is
    \begin{equation}
      E(s) = \frac{\Gamma (k,k s)}{\Gamma (k)}
    \end{equation}
    the mean square free path is
    \begin{equation}
      \langle s^2 \rangle = \frac{1}{k}+1
    \end{equation}
    and the mean square extinction is
    \begin{equation}
      \int_0^\infty E(s) s^2 = \frac{(k+1) (k+2)}{3 k^2}.
    \end{equation}
    In addition to recent time-resolved Gamma-flight solutions~\cite{caer11,pogorui11,pogorui13} we consider the steady-state collision densities of all orders, as well as the scalar flux, and diffusion approximations to both.

  \subsection{Gamma Random Flights---Rod Model ($d = 1$)}
    \subsubsection{Rod Model - Gamma $k = 1/2$}
      For $k = 1/2$, the transformed propagator in the rod is
      \begin{equation}
        \o{\zeta_1}(z) = \frac{\sqrt{z} \sqrt{\frac{\sqrt{4 z^2+1}+1}{4 z^3+z}}}{\sqrt{2}}.
      \end{equation}
      The characteristic equation yields two positive discrete eigenvalues,
      \begin{equation}
        \nu_0 = \frac{\sqrt{\frac{4
   c^2}{c^2-1}-\frac{8}{c^2-1}+\frac{c^4}{c^2-1} \pm \frac{\sqrt{c^2+8}
   c^3}{c^2-1}}}{\sqrt{2}}.
      \end{equation}
      The 2nd collision density can be found explicitly,
      \begin{equation}
        \Cpt(r|2) = \frac{c e^{-r/2} \left(2 e^{r/2} K_0\left(\frac{r}{2}\right)+\pi \right)}{8 \pi}, 
      \end{equation}
      as well as the 4th,
      \begin{equation}
        \Cpt(r|4) = \frac{c^3 \left(\pi  e^{-r/2} (r+6)+8 r K_1\left(\frac{r}{2}\right)\right)}{64\pi }. 
      \end{equation}

    \subsubsection{Rod Model - Gamma $k = 2$}
      For $k = 2$, the transformed Gamma propagator in the rod is
      \begin{equation}
        \o{\zeta_1}(z) = -\frac{4 \left(z^2-4\right)}{\left(z^2+4\right)^2}.
      \end{equation}
      The low order collision densities can be solved for explicitly,
      \begin{align}
        &\Cpt(r|2) = \frac{1}{12} c e^{-2 r} \left(8 r^3+6 r+3\right) \\
        &\Cpt(r|3) = \frac{1}{240} c^2 e^{-2 r} \left(2 r \left(r \left(8 r^3+30
   r+45\right)+45\right)+45\right)
      \end{align}
      Similar to classical transport in the rod, the rigorous collision-density diffusion solution is the exact solution in the rod, but with \emph{two} discrete diffusion terms
      \begin{equation}
        \Cpt(r) = \sum_{\chi} -\frac{\left(\chi ^4-16\right) e^{-r \chi }}{2 c \chi  \left(\chi ^2+12\right)}.
      \end{equation}
      The pair of positive discrete inverse diffusion lengths are
      \begin{equation}
        \chi \in \sqrt{2} \sqrt{2 + c \pm \sqrt{c (c+8)}}.
      \end{equation}

  \subsection{Gamma Random Flights---Flatland ($d = 2$)}
    \subsubsection{Flatland - Gamma $k = 2$}
      For $k = 2$, the transformed Gamma propagator in flatland is
      \begin{equation}
        \o{\zeta_2}(z) = \frac{8}{\left(z^2+4\right)^{3/2}}.
      \end{equation}
      The characteristic equation yields the single positive eigenvalue
      \begin{equation}
        \nu_0 = \frac{1}{2 \sqrt{1-c^{2/3}}}.
      \end{equation}
      The $n$th collision densities can be found explicitly
      \begin{equation}
        \Cpt(r|n) = \frac{2 c^{n-1} r^{\frac{3 n}{2}-1} K_{1-\frac{3 n}{2}}(2 r)}{\pi  \Gamma \left(\frac{3
   n}{2}\right)}.
      \end{equation}
      The $n$th scalar fluxes / fluences can also be found explicitly
      \begin{multline}
        \phipt(r|n) = \frac{c^n r^{\frac{3 n}{2}-\frac{1}{2}}}{4 \Gamma \left(\frac{3
   (n+1)}{2}\right)} [ 2 r I_{-\frac{3}{2} (n-1)}(2 r)-6 n I_{\frac{1}{2} (1-3 n)}(2 r) \\
   + (3 n+1) I_{\frac{1}{2} (3 n-1)}(2 r)-2 r I_{\frac{1}{2} (3 n+1)}(2 r) ] \sec \left(\frac{3 \pi  n}{2}\right).
      \end{multline}

  \subsection{Gamma Random Flights---3D}
    \subsubsection{3D - Gamma $k = 2$}
      For $k = 2$, the transformed propagator in 3D is
      \begin{equation}
        \o{\zeta_2}(z) = \frac{4}{z^2+4}.
      \end{equation}
      The characteristic equation yields the single positive eigenvalue
      \begin{equation}
        \nu_0 = \frac{1}{2 \sqrt{1-c}}.
      \end{equation}
      The $n$th collision densities can be found explicitly
      \begin{equation}
        \Cpt(r|n) = \frac{2 (c r)^{n-1} K_{\frac{3}{2}-n}(2 r)}{\pi ^{3/2} \sqrt{r} \Gamma (n)}.
      \end{equation}
      The exact total collision density is also found to be
      \begin{equation}
        \Cpt(r) = \frac{e^{-2 \sqrt{1-c} r}}{\pi  r}.
      \end{equation}
      Thus, we see a random flight model in 3D with the same spectrum as classical transport in a rod (scaled by a factor of 2), and the exact total collision density is diffusion.  However, we were unable to invert the scalar flux / fluence analytically.

  \subsection{Gamma Random Flights---General Dimension}
    The transformed propagator in $d$ dimensions is
    \begin{equation}\label{eq:gammaprop}
      \zeta_d(z) = \F \{ p(r) / \Omega_d(r) \} = \, _2F_1\left(\frac{k}{2},\frac{k+1}{2};\frac{d}{2};-\frac{z^2}{k^2}\right)
    \end{equation}
    which for the first few integer dimensions gives
    \begin{equation*}
     \begin{array}{llll}
     d & \zeta_d(z), k=2 & \zeta_d(z),k = 3 & \zeta_d(z),k = 4 \\
 1 & -\frac{4 \left(z^2-4\right)}{\left(z^2+4\right)^2} & -\frac{243
   \left(z^2-3\right)}{\left(z^2+9\right)^3} & \frac{256 \left(z^4-96
   z^2+256\right)}{\left(z^2+16\right)^4} \\
 2 & \frac{8}{\left(z^2+4\right)^{3/2}} & -\frac{27 \left(z^2-18\right)}{2
   \left(z^2+9\right)^{5/2}} & -\frac{512 \left(3 z^2-32\right)}{\left(z^2+16\right)^{7/2}}
   \\
 3 & \frac{4}{z^2+4} & \frac{81}{\left(z^2+9\right)^2} & -\frac{256 \left(z^2-48\right)}{3
   \left(z^2+16\right)^3} \\
 4 & \frac{8-\frac{16}{\sqrt{z^2+4}}}{z^2} & \frac{27}{\left(z^2+9\right)^{3/2}} &
   \frac{1}{\left(\frac{z^2}{16}+1\right)^{5/2}} \\
 5 & \frac{12 \left(z-2 \tan ^{-1}\left(\frac{z}{2}\right)\right)}{z^3} & \frac{81
   \left(\left(z^2+9\right) \tan ^{-1}\left(\frac{z}{3}\right)-3 z\right)}{2 z^3
   \left(z^2+9\right)} & \frac{256}{\left(z^2+16\right)^2} \\
    \end{array}
    \end{equation*}

    For Gamma random flights in $d$ dimensions, the first few even collision density moments are
      \begin{align}
        &\int_0^\infty \Omega_d(r) \Cpt(r|n) dr = c^{n-1} \\
        &\int_0^\infty \Omega_d(r) r^2 \Cpt(r|n) dr = \frac{(k+1) n c^{n-1}}{k} \\
        &\int_0^\infty \Omega_d(r) r^4 \Cpt(r|n) dr = \\ &\frac{(k+1) n c^{n-1} (d (k (k n+n+4)+6)+2 k (k+1) (n-1))}{d k^3}
      \end{align}
      and the total collision density moments
      \begin{align}
        &\int_0^\infty \Omega_d(r) \Cpt(r) dr = \frac{1}{1-c} \\
        &\int_0^\infty \Omega_d(r) r^2 \Cpt(r) dr = \frac{k+1}{(c-1)^2 k} \\
        &\int_0^\infty \Omega_d(r) r^4 \Cpt(r) dr = \\ &-\frac{(k+1) (d (c ((k-3) k-6)+(k+2) (k+3))+4 c k (k+1))}{(c-1)^3 d k^3}.
      \end{align}

      The first few even $n$th scattered scalar flux / fluence moments are
      \begin{align}
        &\int_0^\infty \Omega_d(r) \phipt(r|n) dr = c^n \\
        &\int_0^\infty \Omega_d(r) r^2 \phipt(r|n) dr = \frac{(k+1) c^n (3 k n+k+2)}{3 k^2}
      \end{align}
      and the total scalar flux / fluence moments are
      \begin{align}
        &\int_0^\infty \Omega_d(r) \phipt(r) dr = \frac{1}{1-c} \\
        &\int_0^\infty \Omega_d(r) r^2 \phipt(r) dr = \frac{(k+1) (2 c (k-1)+k+2)}{3 (c-1)^2 k^2}.
      \end{align}
      Again, we see that the first two even moments are dimension independent.

    \subsubsection{Gamma Flights Collision Density Diffusion}
      The \changed{Pad\'{e}} approximant producing the transformed classical $P_1$ diffusion approximation for the total collision density is
      \begin{equation}
        \frac{\o{\zeta}_d(z)}{1 - \ssalbedo \o{\zeta}_d(z)} \approx -\frac{2 d \,k}{2 (c-1) d k-(k+2) z^2}
      \end{equation}
      yielding the collision density $P_1$ diffusion length
      \begin{equation}
        \nu_0^2 = \frac{k+2}{2 d k-2 c \, d \, k}.
      \end{equation}
      Likewise, the Grosjean collision density diffusion length is given by
      \begin{equation}
        \nu_G^2 = \frac{(c-2) (k+1)}{2 (c-1) d \, k}.
      \end{equation}

    \subsubsection{Gamma Flights Scalar Flux Diffusion}
      The transformed stretched extinction propagator is
      \begin{equation}
        \o{X(z)} = \, _3F_2\left(\frac{1}{2},\frac{k}{4}+\frac{1}{2},\frac{k}{4}+1;\frac{3}{2},\frac{d}{2};-\frac{4 z^2}{k^2}\right)
      \end{equation}
      where $_3F_2$ is a hypergeometric function.  This yields the scalar flux diffusion length
      \begin{equation}
        \nu_0^2 = \frac{(-k-1) (2 c (k-1)+k+2)}{6 (c-1) d \, k^2}
      \end{equation}
      where agrees with the classical exponential result when $k = 2$.  The Grosjean diffusion length for the scalar flux / fluence is
      \begin{equation}
        \nu_G^2 = \frac{(k+1) (c (k+2)-4 k-2)}{6 (c-1) d \, k^2}.
      \end{equation}

 \section{Pearson (Delta) Random Flights}
  	Here we consider non-classical transport with a Dirac-delta free-path distribution with unit mean,
	\begin{equation}
	  p(s) = \delta(s-1).
	\end{equation}
	This corresponds to the classic Pearson~\shortcite{pearson05} random walk/flight and in some sense is the most extremely correlated generalization of Boltzmann transport theory that we could consider.  The extinction function for Pearson flights is
	\begin{equation}
	  E(s) = 1-\Theta (s-1)
	\end{equation}
	where $\Theta$ is the Heaviside theta function.
	The mean square free path is
	\begin{equation}
	  \langle s^2 \rangle = 1
	\end{equation}
	and the mean square extinction is
	\begin{equation}
	  \int_0^\infty E(s) s^2 = \frac{1}{3}.
	\end{equation}

  \subsection{Pearson Random Flights---Rod Model ($d = 1$)}
      The transformed Pearson propagator in the rod is
      \begin{equation}
        \o{\zeta_1}(z) = \cos (z).
      \end{equation}
      The characteristic equation yields one positive discrete eigenvalue,
      \begin{equation}
        \nu_0 = \frac{1}{\cosh ^{-1}\left(\frac{1}{c}\right)}.
      \end{equation}
      We were unable to derive exact collision densities, but were able to find exact scalar flux / fluence for low order collision counts,
      \begin{align}
      	\phipt(r|2) &= \frac{1}{16} c^2 (\text{sgn}(1-r)+\text{sgn}(3-r)+2) \\
      	\phipt(r|3) &= \frac{1}{32} c^3 (2 \text{sgn}(2-r)+\text{sgn}(4-r)+3) \\
      	\phipt(r|4) &= \frac{1}{64} c^4 (2 \text{sgn}(1-r)+3 \text{sgn}(3-r)+\text{sgn}(5-r)+6) \\
      	\phipt(r|5) &= \frac{1}{128} c^5 (5 \text{sgn}(2-r)+4 \text{sgn}(4-r)+\text{sgn}(6-r)+10)
      \end{align}
      where $\text{sgn}(x)$ is the sign of x,
      \begin{equation}
      	\text{sgn}(x) = \begin{cases} 1 & x > 0 \\ 0 & x = 0 \\ -1 & x < 0. \end{cases}
      \end{equation}
      The rigorous asymptotic diffusion approximation for Pearson flights in a rod is
      \begin{equation}
      	\Cpt(r) \approx \frac{e^{-r \, \text{sech}^{-1}(c)}}{c \sqrt{1-c^2}}.
      \end{equation}

  \subsection{Pearson Random Flights---Flatland ($d = 2$)}

\begin{figure*}
        \centering
        \subfigure[Low absorption]{\includegraphics[width=.8\linewidth]{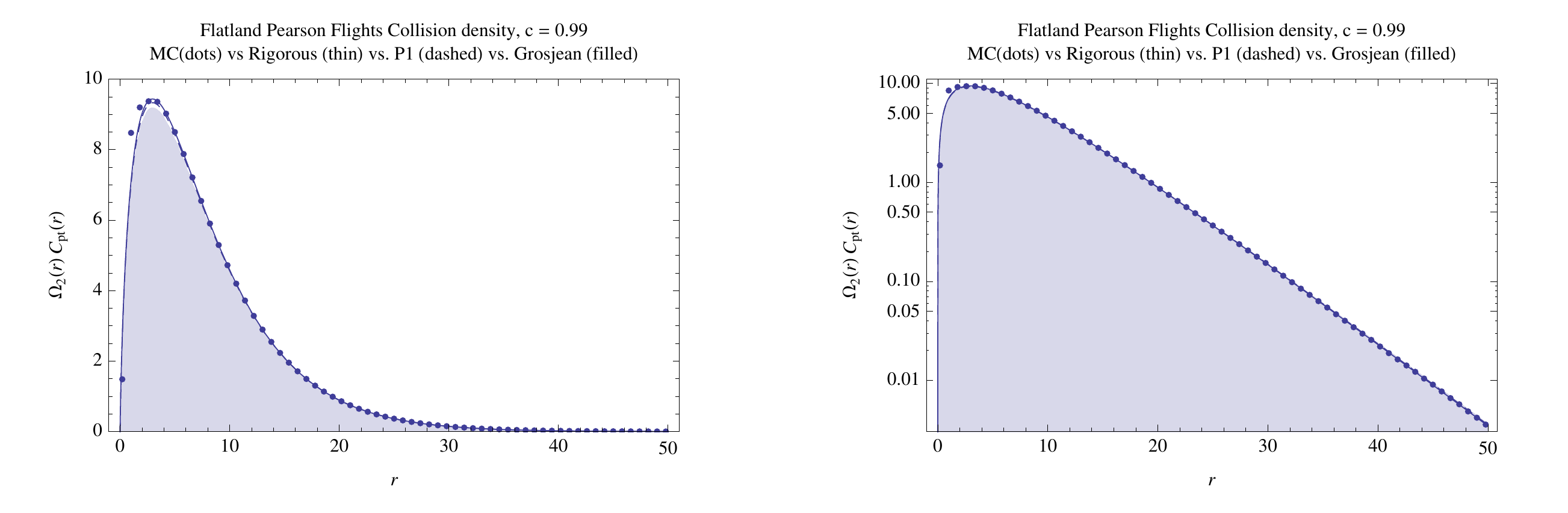}}
        \subfigure[High absorption]{\includegraphics[width=.8\linewidth]{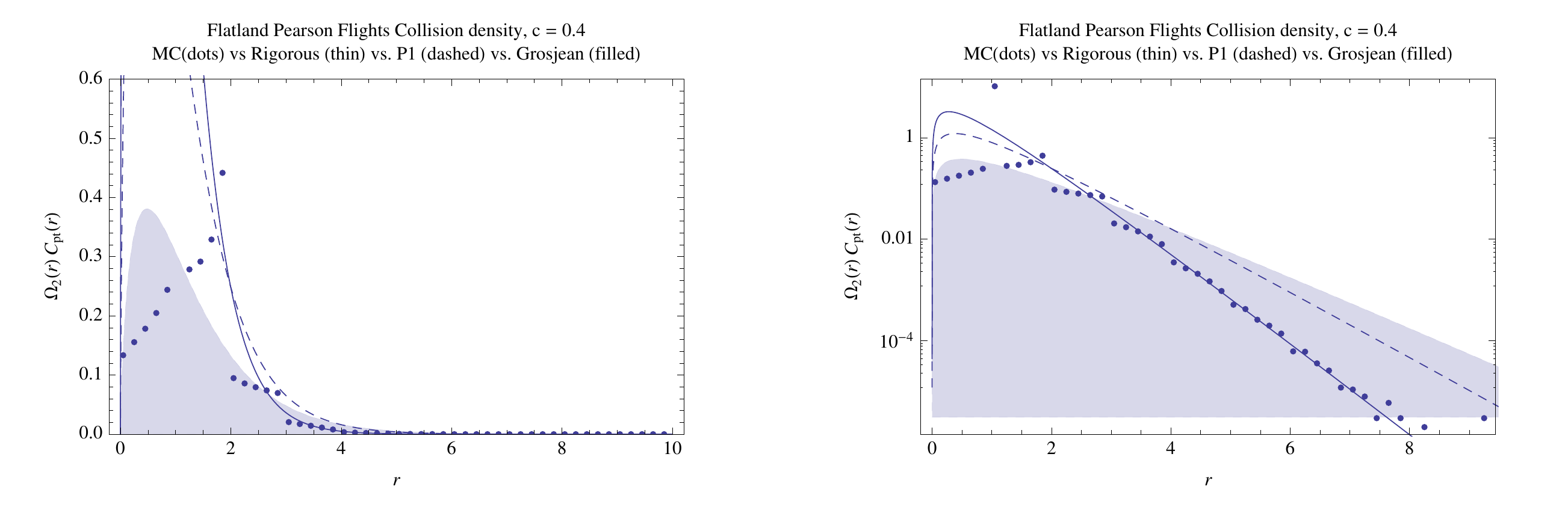}}
        \caption{Comparison of the accuracy of classical ($P_1$), rigorous asymptotic and Grosjean modified diffusion approximations for the collisions density $\Cpt(r)$ about an isotropic point source in an infinite medium in flatland with a Pearson free path distribution.  \changed{Because the Pearson random flight has a deterministic unit step length after every scattering event, the support of each individual scattering order is finite and the total solution is non-smooth.  However, both modified forms of diffusion approximation show improvement over the classical form (Grosjean's method, near the point source, and rigorous far from it).}}
        \label{fig-flatlandPearson} 
      \end{figure*}

    The transformed propagator for Pearson flights in flatland is
      \begin{equation}
        \o{\zeta_2}(z) = J_0(z).
      \end{equation}
      The characteristic equation yields one positive discrete eigenvalue, the solution of
      \begin{equation}
        1 - c I_0(\frac{1}{\nu_0}) = 0,
      \end{equation}
      and the rigorous asymptotic diffusion approximation is
      \begin{equation}
      	\Cpt(r) \approx \frac{K_0\left(\frac{r}{\nu _0}\right)}{\pi  c^2 \nu _0 I_1\left(\frac{1}{\nu _0}\right)}.
      \end{equation}
      In Figure~\ref{fig-flatlandPearson} we compare the rigorous asymptotic diffusion approximation to the classical $P_1$ diffusion approximation for the collision density
      \begin{equation}
      	\Cpt(r) \approx \frac{2 K_0\left(2 \sqrt{1-c} r\right)}{\pi }
      \end{equation}
      and to the Grosjean modified diffusion approximation
      \begin{equation}
      	\Cpt(r) \approx \frac{\delta(r-1)}{2 \pi r} +\frac{2 c K_0\left(2 \sqrt{1+\frac{1}{c-2}} r\right)}{\pi  (2-c)}.
      \end{equation}
      The first few collision densities are known in closed form,
      \begin{equation}
      	\Cpt(r|2) = \begin{cases}
      		\frac{c}{\pi ^2 r \sqrt{4-r^2}} & r<2 \\
      		0 & r \ge 2,
      	\end{cases}
      \end{equation}
      with higher order terms becoming more complex~\cite{borwein11}
	  \begin{equation}
	  	\Cpt(r|3) = c^2 \begin{cases}
	  	  \frac{2 \sqrt{3} r \, _2F_1\left(\frac{1}{3},\frac{2}{3};1;\frac{r^2 \left(9-r^2\right)^2}{\left(r^2+3\right)^3}\right)}{\pi  \left(r^2+3\right)} & r < 3 \\
	       0 & r \ge 3. \end{cases}
	  \end{equation}

  \subsection{Pearson Random Flights---3D ($d = 3$)}

    Pearson flights in 3D were explored in detail by Watson~\shortcite{watson62}.  The transformed propagator for Pearson flights in 3D is
      \begin{equation}
        \o{\zeta_2}(z) = \frac{ \sin z}{z}.
      \end{equation}
      The characteristic equation yields one positive discrete eigenvalue, the solution of
      \begin{equation}
      	1-c \nu_0 \sinh \left(\frac{1}{\nu_0}\right) = 0.
      \end{equation}
      Exact low-order collision densities are known,
      \begin{align}
      	\Cpt(r|2) &= \frac{c (\text{sgn}(2-r)+1)}{16 \pi  r} \\
      	\Cpt(r|3) &= \frac{c^2 (\left| r-3\right| -3 \left| r-1\right| +2 r)}{32 \pi  r} \\
      	\Cpt(r|4) &= \frac{c^3 (-(r-4) \left| r-4\right| +4 (r-2) \left| r-2\right| +(8-3 r) r)}{128 \pi  r}
      \end{align}
      The classical $P_1$ diffusion approximation for the collision density is
      \begin{equation}
      	\Cpt(r) \approx \frac{3 e^{-\sqrt{6-6 c} r}}{2 \pi  r},
      \end{equation}
      the Grosjean modified diffusion approximation is
      \begin{equation}
      	\Cpt(r) \approx \frac{\delta (r-1)}{4 \pi  r^2}+\frac{3 c e^{-\sqrt{\frac{6}{c-2}+6} r}}{2 \pi  (2-c) r},
      \end{equation}
      and the rigorous asymptotic diffusion approximation is
      \begin{equation}
      	\Cpt(r) \approx -\frac{e^{-\frac{r}{\nu _0}}}{\nu _0^2 2 \pi  c r \left(1-\sqrt{c^2+\frac{1}{\nu_0^2}}\right)}.
      \end{equation}
      The transformed stretched extinction propagator for Pearson flights in 3D is
      \begin{equation}
        \o{X}(z) = \frac{\text{Si}(z)}{z}
      \end{equation}
      where Si is the sin integral function
      \begin{equation}
      	\text{Si}(z) = \int_0^z \frac{\sin t}{t} dt.
      \end{equation}
      The classical $P_1$ diffusion approximation for the scalar flux / fluence is thus
      \begin{equation}
      	\phipt(r) \approx \frac{9 e^{-3 \sqrt{\frac{3}{2 c+1}-1} r}}{\pi  (4 c+2) r}
      \end{equation}
      and the Grosjean diffusion approximation is
      \begin{equation}
      	\phipt(r) \approx \frac{E(r)}{4 \pi r^2} + \frac{9 c e^{-3 \sqrt{\frac{6}{c-4}+2} r}}{2 \pi  (4-c) r}.
      \end{equation}

  \subsection{Pearson Random Flights---General Dimension}
    The transformed Pearson propagator in $d$ dimensions is
    \begin{multline}
      \zeta_d(z) = \F \{ p(r) / \Omega_d(r) \} \\ = \frac{2^{d/2} z^{1-\frac{d}{2}} \Gamma \left(\frac{d}{2}+1\right) J_{\frac{d}{2}-1}(z)}{d} 
      = \, _0F_1\left(;\frac{d}{2};-\frac{z^2}{4}\right)
    \end{multline}
    and the transformed stretched extinction propagator is
    \begin{equation}
    	\o{X}(z) = \, _1F_2\left(\frac{1}{2};\frac{3}{2},\frac{d}{2};-\frac{z^2}{4}\right)
    \end{equation}
    where $_1F_2$ is a hypergeometric function.

    For Gamma random flights in $d$ dimensions, the first few even collision density moments are
      \begin{align}
        &\int_0^\infty \Omega_d(r) \Cpt(r|n) dr = c^{n-1} \\
        &\int_0^\infty \Omega_d(r) r^2 \Cpt(r|n) dr = n c^{n-1} \\
        &\int_0^\infty \Omega_d(r) r^4 \Cpt(r|n) dr = \frac{n c^{n-1} ((d+2) n-2)}{d} \\
        &\int_0^\infty \Omega_d(r) r^6 \Cpt(r|n) dr = \frac{n c^{n-1} ((d+4) n ((d+2) n-6)+16)}{d^2}
       \end{align}
      and the total collision density moments
      \begin{align}
        &\int_0^\infty \Omega_d(r) \Cpt(r) dr = \frac{1}{1-c} \\
        &\int_0^\infty \Omega_d(r) r^2 \Cpt(r) dr = \frac{1}{(1-c)^2} \\
        &\int_0^\infty \Omega_d(r) r^4 \Cpt(r) dr = -\frac{c (d+4)+d}{(c-1)^3 d} \\
        &\int_0^\infty \Omega_d(r) r^6 \Cpt(r) dr = \frac{c^2 (d (d+12)+48)+4 c d (d+6)+d^2}{(c-1)^4 d^2}.
      \end{align}

      The first few even $n$th scattered scalar flux / fluence moments are
      \begin{align}
        &\int_0^\infty \Omega_d(r) \phipt(r|n) dr = c^n \\
        &\int_0^\infty \Omega_d(r) r^2 \phipt(r|n) dr = \frac{1}{3} (3 n+1) c^n \\
        &\int_0^\infty \Omega_d(r) r^4 \phipt(r|n) dr = \frac{c^n \left(15 (d+2) n^2+10 (d-1) n+3 d\right)}{15 d}
      \end{align}
      and the first two even scalar flux / fluence moments are
      \begin{align}
        &\int_0^\infty \Omega_d(r) \phipt(r) dr = \frac{1}{1-c} \\
        &\int_0^\infty \Omega_d(r) r^2 \phipt(r) dr = \frac{2 c+1}{3 (c-1)^2}.
      \end{align}

    \subsubsection{Pearson Flights Collision Density Diffusion}
      The \changed{Pad\'{e}} approximant producing the transformed classical $P_1$ diffusion approximation for the total collision density is
      \begin{equation}
        \frac{\o{\zeta}_d(z)}{1 - \ssalbedo \o{\zeta}_d(z)} \approx \frac{2 d}{z^2-2 (c-1) d}
      \end{equation}
      yielding the collision density $P_1$ diffusion length
      \begin{equation}
        \nu_0^2 = \frac{1}{2(1-c)d}.
      \end{equation}
      Likewise, the Grosjean collision density diffusion length is given by
      \begin{equation}
        \nu_G^2 = \frac{2-c}{2 (1-c) d}.
      \end{equation}

    \subsubsection{Pearson Flights Scalar Flux Diffusion}
      The transformed stretched extinction propagator is
      \begin{equation}
        \o{X(z)} = \, _1F_2\left(\frac{1}{2};\frac{3}{2},\frac{d}{2};-\frac{z^2}{4}\right)
      \end{equation}
      where $_1F_2$ is a hypergeometric function.  This yields the scalar flux diffusion length
      \begin{equation}
        \nu_0^2 = -\frac{2 c+1}{6 (c-1) d}.
      \end{equation}
      The Grosjean diffusion length for the scalar flux / fluence is
      \begin{equation}
        \nu_G^2 = \frac{c-4}{6 (c-1) d}.
      \end{equation}

\section{Hand-crafted transport theory}
	We are free to invert the spectrum of exponential scattering in a given dimension $d$, but perform the inversion into another dimensional space $d'$, to find a free-path distribution such that the spectrum of the transport operator is the familiar spectrum from classical transport in the original dimension $d$.  For example, starting with the diffusion spectrum for exponential scattering in the Rod model (Equation~\ref{eq:exprodprop}), we find that the free-path distribution required to make diffusion an exact solution in \changedcomment{wording}{a $d$-dimensional space} is
	\begin{equation}
		p(s) = \Omega_d(r) \Finv \{ \frac{1}{1 + z^2}\}= \frac{s^{d/2} 2^{-d/2} d \, K_{\frac{d-2}{2}}(s)}{\Gamma \left(\frac{d}{2}+1\right)}.
	\end{equation}
	So for the total collision density in flatland to be exactly diffusion, the free-path distribution must be
	\begin{equation}
		p(s) = s K_0(s)
	\end{equation}
	or some scaled version thereof.  Similarly, for diffusion to be exact in 3D, we have already seen the distribution must be a Gamma distribution
	\begin{equation}
		p(s) = e^{-s} s,
	\end{equation}
	and so on.  To produce the flatland exponential spectrum in the Rod model, the free-path distribution must be
	\begin{equation}
		p(s) = \frac{2 K_0(s)}{\pi }
	\end{equation}
	and to produce the 3D exponential spectrum in the Rod model, the free-path distribution must be
	\begin{equation}
		p(s) = -\text{Ei}(-s).
	\end{equation}
	However, this does not extend generally---we were unable to find the free-path distribution in 3D such that the spectrum of the transport operator had the simple $\sqrt{1-c^2}$ eigenvalue seen in flatland.

	We have also seen the Chi free-path distributions with $k = d$ create Gaussian per-order collision densities for all orders $n$.

\section{Discussion and Conclusions}

	We have generalized two non-classical approaches to forming diffusion approximations for steady-state monoenergetic Boltzmann transport in infinite media.  Both of these approaches differ significantly from classical $P_1$ diffusion and provide higher accuracy at relatively little cost.  Grosjean's modified diffusion theory provides improved accuracy for high absorption levels and for small source-detector separations and rigorous asymptotic diffusion provides improved accuracy for very low densities far from sources.  We have extended both of these diffusion theories by considering generalized linear-Boltzmann transport processes with non-exponential free-path distributions and, further, by considering diffusion approximations in arbitrary dimensional spaces.

	We found that Grosjean's modified diffusion easily provides improved accuracy for exponential transport in arbitrary dimensions.  We further demonstrated its utility for several families of non-exponential free-path distributions.  For the case of non-exponential free-paths, the total collision density and the scalar flux / fluence are no longer proportional distributions and we found differing diffusion lengths and approximations for each.

	By identifying poles of the Fourier-transformed total collision density and scalar flux / fluence, we were successfully able to derive accurate rigorous asymptotic diffusion approximations for a variety of transport types in various dimensions.  As is the case for classical transport in 3D, we expect these rigorous discrete diffusion terms to reflect the discrete spectrum of the generalized transport operator in other dimensions and for non-exponential free-path distributions.  In deriving these rigorous asymptotic diffusion modes we saw some unfamiliar behaviours arise as we departed from classical transport in 3D.  For exponential transport in 6D with isotropic scattering we found a pair of discrete eigenvalues for the transport operator, creating a sum of two diffusion terms.  Thus, high dimensionality changes the structure of the exact solution of an isotropically-scattered process in the same way that changing the scattering kernel in a fixed dimensional space does.  Additionally, we saw the breakdown of the rigorous asymptotic diffusion approximation in 4D and 6D when the absorption level was too high.  We found transport theories where diffusion is no longer an exact solution in the rod model, a transport theory where the exact solution in the rod model is a sum of two diffusion terms, as well as discovering transport theories where diffusion is an exact solution in higher dimensions.

	We interpreted the derivation of diffusion approximations as inverse Fourier transforms of frequency-domain \changed{Pad\'{e}} approximants.  This allowed broad validation of a conjecture that both $P_1$ and Grosjean diffusion lengths are purely a function of the mean free path, the mean square free path, and the dimension $d$.  This also produced a proof of the $1 / \sqrt{d}$ dependence on the \changedcomment{specify which diffusion, since this does not apply to the rigorous}{$P_1$ and Grosjean} diffusion length\changed{s} in any dimension $d$ for exponential random flights, and we further demonstrated this behaviour for several families of non-exponential free-path distributions.

\section{Acknowledgements}
	The author is indebted to MMR Williams, Norm McCormick, and Barry Ganapol for answering many questions about transport theory \changed{and to Jean-Marie Aubry for a helpful discussion regarding residue theory}.  Further, we wish to thank Alwin Kienle for clarifying the derivation of his result for exponential scattering in flatland.

\bibliographystyle{acmsiggraph}
\bibliography{flatland}

\end{document}